\documentclass[12pt,floatfix,aps,nofootinbib,superscriptaddress]{revtex4} 
\pdfoutput=1 
\usepackage{amssymb,amsfonts,multirow}

\usepackage{epsfig}

\usepackage{placeins}
\usepackage{xspace}
\usepackage{cancel}

\usepackage{amssymb,url}
\usepackage{graphicx}
\usepackage{hyperref}

\usepackage{color}

\usepackage{array}
\usepackage{amsmath}
\usepackage{slashed}

\definecolor{rossoCP3}{cmyk}{0,.88,.77,.40}

\long\def\del #1 \enddel { }

\usepackage{graphicx}
\usepackage{amsmath}
\usepackage{amssymb}
\usepackage{subfigure}

\usepackage{epsfig}

\usepackage{graphicx}
\usepackage{subfigure}
\usepackage{hyperref}

\def\beq{\begin{equation}}
\def\eeq{\end{equation}}

\def\bea{\arraycolsep .1em \begin{eqnarray}}
\def\eea{\end{eqnarray}}
\def\Tr{{\rm Tr}}

\def\eps{\epsilon}

\def\al#1{\alpha_{#1}}
\def\eq#1{(\ref{#1})}

\def\s0#1#2{\mbox{\small{$ \frac{#1}{#2} $}}}
\def\0#1#2{\frac{#1}{#2}}

\def\grgl{\:\hbox to -0.2pt{\lower2.5pt\hbox{$\sim$}\hss}{\raise3pt\hbox{$>$}}\:}
\def\klgl{\:\hbox to -0.2pt{\lower2.5pt\hbox{$\sim$}\hss}{\raise3pt\hbox{$<$}}\:}

\def\lsim{\mathrel{\rlap{\lower4pt\hbox{\hskip1pt$\sim$}}
    \raise1pt\hbox{$<$}}}                
\def\gsim{\mathrel{\rlap{\lower4pt\hbox{\hskip1pt$\sim$}}
    \raise1pt\hbox{$>$}}}                

\begin{document}
${}$\vskip1cm

\title{Asymptotic safety guaranteed}
\author{Daniel~F.~Litim}
\email{d.litim@sussex.ac.uk}
\affiliation{\mbox{Department of Physics and Astronomy, U Sussex, Brighton, BN1 9QH, U.K.}}
\author{Francesco~Sannino}
\email{sannino@cp3-origins.net}
\affiliation{{\color{rossoCP3}CP${}^3$-Origins} \& the Danish Institute for Advanced Study Danish IAS,
Univ. of Southern Denmark, Campusvej 55, DK-5230 Odense}

\begin{abstract}
We study the ultraviolet behaviour of four-dimensional quantum field theories involving non-abelian gauge fields, fermions and scalars in the Veneziano limit.
In a regime where asymptotic freedom is lost, we explain how the three types of fields cooperate to develop fully interacting  
ultraviolet fixed points, strictly controlled by perturbation theory. 
Extensions towards strong coupling and beyond the large-$N$ limit are discussed.
\vskip10.7cm
{\noindent \footnotesize Preprint: CP3-Origins-2014-24, DNRF90 \& DIAS-2014-24}

\end{abstract}

\maketitle
\newpage
\tableofcontents

\section{Introduction}

It is widely acknowledged that ultraviolet (UV) fixed points are central for quantum field theories  to be fundamental and predictive  up  to highest energies \cite{Wilson:1971bg,Wilson:1971dh}.
 A well-known example 
 is asymptotic freedom of quantum chromodynamics where the UV fixed point 
is non-interacting
 \cite{Gross:1973id,Politzer:1973fx}. In turn, neither the $U(1)$ nor the scalar sector of the standard model are asymptotically free. This is known as the triviality problem, which limits the predictivity  to a scale of maximal  UV extension \cite{Callaway:1988ya}.  
High-energy fixed points may also be interacting, a scenario referred to as asymptotic safety \cite{Weinberg:1980gg}.  It is then tempting to think   
that theories  which are not asymptotically free,
or not even  renormalisable by power-counting,
may well turn out to be fundamental in their own right, provided
they develop an interacting UV fixed point  \cite{Litim:2011cp}.
In recent years, asymptotic safety has become a popular scenario to address quantum aspects of  gravity  \cite{Litim:2011cp,Litim:2006dx,Niedermaier:2006ns,Niedermaier:2006wt,
Percacci:2007sz,Litim:2008tt,Reuter:2012id}. In a similar vein, UV conformal extensions of the standard model with and without gravity have received some attention in view of interacting fixed points 
\cite{Kazakov:2002jd,Gies:2003dp,Morris:2004mg,Fischer:2006fz,
Fischer:2006at,Kazakov:2007su,Zanusso:2009bs,
Gies:2009sv,Daum:2009dn,Vacca:2010mj,Calmet:2010ze,Folkerts:2011jz,
Bazzocchi:2011vr,Gies:2013pma,Antipin:2013exa,Dona:2013qba} 
and 
scale invariance in particle physics and cosmology \cite{Bonanno:2001xi,Meissner:2006zh,Foot:2007iy,Hewett:2007st,Litim:2007iu,Shaposhnikov:2008xi,
Shaposhnikov:2008xb,Shaposhnikov:2009pv,Weinberg:2009wa,
Hooft:2010ac,Gerwick:2011jw,Gerwick:2010kq,
Hindmarsh:2011hx,Hur:2011sv,Dobrich:2012nv,Tavares:2013dga,
Tamarit:2013vda,Abel:2013mya,Antipin:2013bya,Heikinheimo:2013fta,
Gabrielli:2013hma,Holthausen:2013ota,Dorsch:2014qpa,Eichhorn:2014qka}.

The most notable difference between asymptotically free and asymptotically safe theories relates to residual interactions at high energies. Canonical power counting becomes modified and the  relevant or marginal invariants which dominate high energy physics  are no longer known a priori.
Couplings may become large and small expansion parameters are often not  available.
Establishing or refuting asymptotic safety  in a reliable manner then becomes a challenging non-perturbative task  \cite{Falls:2013bv}. 
A few  rigorous results for asymptotically safe UV fixed points have been obtained for certain  power-counting non-renormalisable models
by taking the space-time dimensionality as a continuous parameter \cite{Gastmans:1977ad,Christensen:1978sc,Weinberg:1980gg,Peskin:1980ay,
Gawedzki:1985uq,Gawedzki:1985ed,Morris:2004mg}  in the spirit of the $\eps$-expansion \cite{Wilson:1973jj}, or by using large-$N$ techniques  \cite{Tomboulis:1977jk,Tomboulis:1980bs,Smolin:1981rm,
deCalan:1991km,Kazakov:2007su,Antipin:2011ny,Antipin:2011aa,
Antipin:2012kc,Antipin:2013pya}. 
Asymptotic safety then arises in the vicinity of the Gaussian fixed point where perturbation theory
is applicable. The success of well-controlled model studies provides valuable starting points to search for asymptotic safety at strong coupling.

In this paper, we are interested in the UV behaviour of interacting gauge fields, fermions and scalars in four dimensions.
In the regime where asymptotic freedom is lost, we ask the question whether the theory is able, dynamically, to 
develop  an interacting UV fixed point.
Our main tool to answer this question is a suitably chosen large-$N$ limit \cite{Veneziano:1979ec} (where $N$ refers to the number of fields), whereby the theory is brought under strict perturbative control. 
 Banks and Zaks \cite{Banks:1981nn} have used a similar idea to investigate the presence of interacting infrared (IR) fixed points in gauge theories with fermionic matter.
Here, we will  discover that all three types of fields are required for an asymptotically safe UV fixed point  to emerge. Given that exactly solvable models in four dimensions are hard to come by, our findings
are a useful starting point to construct UV safe models of particle physics.

   \begin{figure*}[t]
\begin{center}
\includegraphics[width=.6\hsize]{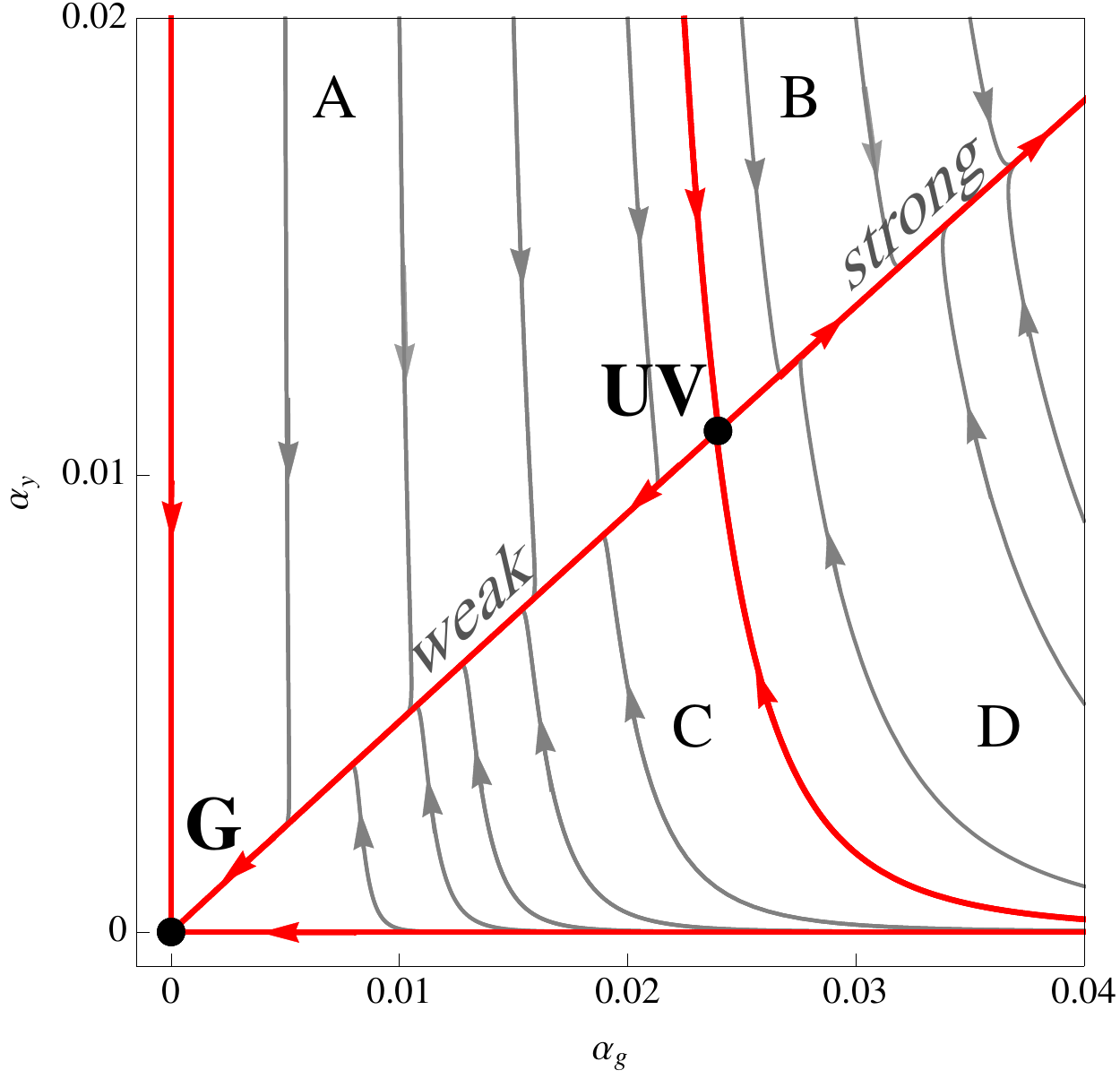}
\caption{\label{pPD}  
The phase diagram of certain 4D gauge-Yukawa theories in the absence of asymptotic freedom  and supersymmetry, in the vicinity of an asymptotically safe fixed point (UV). 
The  renormalization group flow for the gauge coupling $\al g$ and the Yukawa coupling $\al y$ is pointing towards the IR. 
The two renormalization group  trajectories emanating out of the asymptotically safe fixed point  define UV finite quantum field theories which at low energies correspond to a weakly (G) and a strongly coupled  theory. 
Other trajectories exist as well, but they do not lead to UV finite theories (within perturbation theory). Parameter choices and further details are given in 
Sec.~\ref{PD}. } 
 \vskip-.3cm
\label{pPhase}
\end{center}
\end{figure*}

We illustrate the main outcome for gauge-matter theories in the absence of asymptotic freedom
in terms of a running non-Abelian coupling $(\al g)$ and a running Yukawa coupling $(\al y)$. Fig.~\ref{pPhase} shows  the phase diagram close to the Gaussian fixed point to leading order in perturbation theory. Renormalization group trajectories are directed towards the IR. 
We observe that both the Yukawa and the gauge coupling behave QED-like close to the Gaussian. Consequently, not a single trajectory  emanates from the Gaussian, meaning that it is an IR fixed point. 
The main novelty is the occurrence of an interacting fixed point induced by
fluctuations of the gauge, fermion, and scalar fields.  The fixed point is located close to the Gaussian and controlled by perturbation theory. 
The UV nature of the fixed point is evidenced by the (two) UV finite renormalization group trajectories running out of it.   They lead to sensible theories at all scales: in the UV, they are    finite due to the interacting fixed point.  In the IR, they correspond to weakly coupled theories with Gaussian scaling,  and to strongly coupled theories with confinement and chiral symmetry breaking or conformal behaviour, respectively. We also find UV unstable  trajectories which  do not emanate from the UV fixed point. They equally approach sensible IR theories, yet their UV predictivity is limited (at least in perturbation theory) by a scale of maximal UV extension.
In this sense,  asymptotic safety 
guarantees the existence of   UV finite matter-gauge theories
even in the absence of  asymptotic freedom or supersymmetry.

In the rest  of the paper  we provide the details of our study.
We recall the perturbative origin of  interacting UV fixed points and asymptotic safety for sample theories of self-interacting  gravitons, fermions, gluons, and scalars
(Sec.~\ref{mechanisms}). We then explain why and how asymptotic safety can arise for  certain gauge-Yukawa theories  in strictly  four dimensions.
Fully interacting UV fixed points are found 
and analysed together with their universal scaling exponents, the UV critical surface, and the phase diagram of Fig.~\ref{pPD}  (Sec.~\ref{AFAS}).  Aspects such as stability, Weyl consistency, unitarity, and triviality are  discussed (Sec.~\ref{consistency}), together with further directions for asymptotic safety within perturbation theory and beyond (Sec.~\ref{LargeCoupling}). We close with some conclusions (Sec.\ref{discussion}).

\section{Origin of interacting UV fixed points}\label{mechanisms}

In this section, we recall the perturbative origin of asymptotic safety for certain quantum field theories. We discuss key examples and introduce some notation. 

\subsection{Asymptotic safety}
Asymptotic safety is the scenario which generalises the notion of a free, Gaussian, ultraviolet fixed point to an interacting, non-Gaussian one. 
An asymptotivcally safe UV fixed point then acts as an anchor for the renormalisation group evolution of couplings, allowing them to approach the high-energy limit along well-defined RG trajectories without encountering  divergences such as Landau poles.

Within perturbation theory, the origin for asymptotic safety is best illustrated in terms of a running dimensionless coupling $\alpha=\alpha(t)$ of a hypothetical theory (to be specified below) with its renormalisation group (RG) $\beta$-function given by
\begin{equation}\label{dalpha}
\partial_t\, \alpha=A \,\alpha - B\,\alpha^2\,.
\end{equation}
Here, $t=\ln (\mu/\Lambda)$ denotes the logarithmic RG `time', $\mu$ the RG momentum scale and $\Lambda$ a characteristic reference scale of the theory. $A$ and  $B$ are numbers. We assume that \eq{dalpha} arises from a perturbative expansion of the full $\beta$-function $(\beta\equiv\partial_t\alpha)$, with $\alpha$ reasonably small for perturbation theory to be applicable. The linear term  relates to the tree level contribution, reflecting that the underlying coupling is dimensionful with mass dimension $-A$. The quadratic term stands for the one loop contribution.
Evidently, the flow displays two types of fixed points, a trivial one at $\alpha_*=0$, and a non-trivial one at 
\begin{equation}\label{FP}
\alpha_*=A/B\,.
\end{equation}
In the spirit of perturbation theory, the non-trivial fixed point \eq{FP} is accessible in the domain of validity of the RG flow \eq{dalpha} as long as $\alpha_*\ll 1$. This can be achieved in two manners, either by having $A\ll 1$ at fixed $B$, or by making $1/B\ll 1$ at fixed $A$. (Below, we discuss examples where both of these options are realised.) 
Integrating \eq{dalpha} in the vicinity of the fixed point, one finds that small deviations from it scale as
\begin{equation}\label{delta}
\delta\alpha=(\alpha-\alpha_*)\propto \left(\frac{\mu}{\Lambda_c}\right)^\vartheta\,,
\end{equation}
thereby relating the characteristic energy scale $\Lambda_c$ of the theory to the deviation from the fixed point $\delta\alpha$ at the RG scale $\mu$ and a universal number $\vartheta$. The role of $\Lambda_c$ here is similar to that of $\Lambda_{\rm QCD}$ in QCD as it describes the cross-over between two different scaling regimes of the theory. The universal scaling index $\vartheta$  arises as the eigenvalue of the one-dimensional `stability matrix'
\begin{equation}
\vartheta=\left.\frac{\partial\beta}{\partial\alpha}\right|_*\,.
\end{equation}
It is given by $\vartheta=-A$ at the non-trivial fixed point \eq{FP}, and by $\vartheta=A$ at the Gaussian fixed point.\footnote{
The leading exponent $\vartheta$ is related to the exponent $\nu$  in the statistical physics literature as $\nu=-1/\vartheta$.} 
 We have that $\vartheta<0$ at the non-trivial fixed point provided $A>0$. Consequently, small  deviations from the fixed point \eq{delta} decrease with increasing RG momentum scale meaning that  \eq{FP} is an UV fixed point. If, additionally, $B>0$, the fixed point obeys $\alpha_*>0$. Consequently, in this case the Gaussian fixed point of the model \eq{dalpha} becomes an infrared (IR) fixed point. Unlike for asymptotically free theories where the RG running close to a trival UV fixed point is logarithmically slow, the RG running close to a non-trivial fixed point is power-like, and thus much faster. We notice that  for $A=0$ the model \eq{dalpha} displays a doubly-degenerate Gaussian fixed point. In this light, as soon as the canonical mass dimension of the underlying coupling becomes negative, $A>0$, such as in theories which are power-counting non-renormalisable,  the degeneracy of the perturbative $\beta$-function is lifted leading to a pair of non-degenerate fixed points. Provided that $\alpha$ is the sole relevant coupling of the model under consideration, the existence of an interacting UV fixed point can be used to define the theory fundamentally. 
This are the bare bones of asymptotic safety \cite{Wilson:1971bg,Wilson:1971dh,Weinberg:1980gg}.

\subsection{Gravitons}
We now recall specific examples for asymptotically safe quantum field theories where the mechanism just described is at work. We start with Einstein gravity with action $(16\pi\,G_N)^{-1}\int\sqrt{\det g} R$ in $D$ dimensions. Newton's coupling $G_N$ has canonical mass dimension $[G_N]=2-D$ and the theory is power-counting non-renormalisable above its critical dimension $D_c=2$. In units of the RG scale, the dimensionless gravitational coupling of the model reads 
\begin{equation}\label{g}
\alpha=G_N(\mu) \mu^{D-2}\,.
\end{equation}
In $D=D_c+\eps$ dimensions, one finds the RG flow \eq{dalpha} with $A=\eps\ll 1$, $B=50/3$, and an UV fixed point \eq{FP} in the perturbative regime by analytical continuation of space-time dimensionality 
\cite{Gastmans:1977ad,Christensen:1978sc,Weinberg:1980gg,Kawai:1992np}. Results have  been extended to two-loop order, also including a cosmological constant \cite{Kawai:1993mb}. It is the sign of $B>0$ which enables an interacting UV  fixed point for gravity in the perturbative domain. Much of the recent motivation to revisit asymptotic safety for gravity in four dimensions derives from this result close to two dimensions \cite{Litim:2011cp,Litim:2006dx,Niedermaier:2006ns,Niedermaier:2006wt,
Percacci:2007sz,Litim:2008tt,Reuter:2012id}. Advanced non-perturbative studies predict 
a  gravitational fixed point  \eq{g} of order unity \cite{Falls:2013bv}, see also \cite{Litim:2003vp,Codello:2008vh,Benedetti:2009rx,Donkin:2012ud,Christiansen:2012rx,
Christiansen:2014raa,Becker:2014qya}.

\subsection{Fermions}
Next we consider a purely fermionic theory of $N_F$ selfcoupled massless Dirac fermions with, examplarily, Gross-Neveu-type selfinteraction 
$\s012 g_{\rm GN}(\bar\psi\psi)^2$, \cite{Gross:1974jv}.
The quartic fermionic selfcoupling $g_{\rm GN}$ has canonical mass dimension  $[g_{\rm GN}]=2-D$ and the model is perturbatively non-renormalisable above its critical dimension  $D_c=2$. The dimensionless coupling reads 
\begin{equation}\label{gGN}
\alpha=\frac{g_{\rm GN}(\mu)}{2\pi N_F} \mu^{2-D}\,.
\end{equation} 
Close to two dimensions  
 the $\beta$-function \eq{dalpha} for \eq{gGN} can be computed within the $\epsilon$-expansion by setting $D=D_c+\epsilon$. The coefficient $A$, given by minus the canonical mass dimension, becomes $A=\epsilon\ll 1$, while the coefficient $B>0$, to leading order in $\epsilon$, is of order one and given by the 1-loop coefficient in the two-dimensional theory. Hence, the model has a reliable UV fixed point \eq{FP} in the perturbative regime. Its renormalisability has been established more rigorously in \cite{Gawedzki:1985uq,Gawedzki:1985ed} with the help of the non-perturbative renormalisation group. Similarly, in the large-$N_F$ limit and at fixed dimension $D=3$, one finds that $A\propto 1/N_F\ll 1$ while the coefficient $B>0$ remains of order unity, 
leading to the same conclusion \cite{deCalan:1991km}. UV fixed points of four-fermion theories have been studied non-perturbatively in the continuum and on the lattice, e.g.
\cite{Kikukawa:1989fw,He:1991by,Hands:1992be}. For a simple example, see \cite{Braun:2010tt}.

\subsection{Gluons}
Next, we consider pure $SU(N_C)$ Yang-Mills theory with $N_C$ colors in $D=4+\eps$ dimensions, with action $\sim 1/(g^2_{\rm YM})\int \Tr F_{\mu\nu}F^{\mu\nu}\,.$ The canonical mass dimension of the coupling reads $[g^2_{\rm YM}]=4-D$ and the theory is perturbatively non-renormalisable above its critical dimension $D_c=4$. Introducing the running dimensionless strong coupling as
\begin{equation}
\alpha=\frac{g^2_{\rm YM}\,N_C}{(4\pi)^2}\,\mu^{D-4}\,,
\end{equation} 
one may derive its $\beta$-function \eq{dalpha} to leading order in the $\eps$-expansion, $\eps = D-D_c$. Again, one finds that $A=\eps\ll1$ and $B>0$ of order unity, thus leading to an UV fixed point \eq{FP} \cite{Peskin:1980ay}. The expansion has been extended up to fourth order in $\eps$ \cite{Morris:2004mg} suggesting that this fixed point persists in $D=5$, in accord with a prediction using functional renormalisation  \cite{Gies:2003ic}. 

\subsection{Scalars}

Finally we turn to self-interacting scalar fields. For scalar field theory with linearly realised $O(N)$ symmetry, the dimensionless version of its quartic self-coupling is given by
\begin{equation}
\alpha=\frac{\lambda\,N}{(4\pi)^2}\,\mu^{D-4}\,.
\end{equation}  
The critical dimension of these models is $D_c=4$. Within the $\eps$-expansion away from the critical dimension, using $D=4-\eps$, the one loop $\beta$-function is of the form \eq{dalpha} with $A=-\eps<0$ and $B<0$. Therefore, the fixed point \eq{FP} is physical below the critical dimension,
and it is 
an infrared one, i.e.~the seminal Wilson-Fisher fixed point. For the physically relevant dimension $D=3$, its existence has been confirmed even beyond perturbation theory
\cite{Pelissetto:2000ek,Litim:2010tt}. In this light, the Wilson-Fisher fixed point can be viewed as the infrared analogue of asymptotic safety. The search for interacting UV fixed points in $D>4$ dimensions within perturbation theory has seen renewed interest  recently \cite{Fei:2014yja}. Functional renormalisation suggests the absence of global fixed points \cite{Percacci:2014tfa}, in accord with the qualitative picture obtained here.

In non-linear $\sigma$-models, i.e.~scalar theories with non-linearly realised internal symmetry, the critical dimension is reduced to $D_c=2$. The relevant coupling then displays an UV fixed point within perturbation theory \cite{Brezin:1975sq,Bardeen:1976zh}. Fixed points of non-linear $\sigma$-models have also been investigated for their similarity with gravity \cite{Percacci:2009dt}.  Lattice results for an interacting UV fixed point in $D=3$ in accord with functional renormalisation have recently been reported in \cite{Wellegehausen:2014noa}.

\section{From asymptotic freedom to asymptotic safety}\label{AFAS}
In this section, we explain the perturbative origin for asymptotic safety  in a class of gauge-Yukawa theories in  strictly four dimensions.

\subsection{Gauge-Yukawa theory}

We consider a theory with $SU(N_C)$ gauge fields $A^a_\mu$ and field strength $F^a_{\mu\nu}$ $(a=1,\cdots, N_C^2-1)$, $N_F$ flavors of fermions $Q_i$ $(i=1,\cdots,N_F)$ in the fundamental representation, and a $N_F\times N_F$ complex matrix scalar field $H$ uncharged under the gauge group. The fundamental action is taken to be the sum of the Yang-Mills action, the fermion kinetic terms, the Yukawa coupling, and the scalar kinetic and self-interaction Lagrangean $L=L_{\rm YM}+L_F+L_Y+L_H+L_U+L_V$, with
\bea
\label{F2}
L_{\rm YM}&=& - \frac{1}{2} \Tr \,F^{\mu \nu} F_{\mu \nu}\\
\label{F}
L_F&=& \Tr\left(
\overline{Q}\,  i\slashed{D}\, Q \right)
\\
\label{Y}
L_Y&=&
y \,\Tr\left(\overline{Q}_L H Q_R + \overline{Q}_R H^\dagger Q_L\right)
\\
\label{H}
L_H&=&\Tr\,(\partial_\mu H ^\dagger\, \partial^\mu H) \\
\label{U}
L_U&=&
-u\,\Tr\,(H ^\dagger H )^2  \,\\
\label{V}
L_V&=&
-v\,(\Tr\,H ^\dagger H )^2  \,. 
\eea
$\Tr$ is the trace over both color and flavor indices, and the decomposition $Q=Q_L+Q_R$ with $Q_{L/R}=\frac 12(1\pm \gamma_5)Q$ is understood. This theory has been investigated for its interesting properties in \cite{Antipin:2011ny,Antipin:2011aa,Antipin:2012kc,Antipin:2013pya}. 
We will motivate it for our purposes while we progress.
In four dimensions, the model has four classically marginal coupling constants given by the gauge coupling, the Yukawa coupling $y$,  the quartic scalar couplings $u$ and the `double-trace' scalar coupling $v$,  which we write as
\beq\label{couplings}
\al g=\frac{g^2\,N_C}{(4\pi)^2}\,,\quad
\al y=\frac{y^{2}\,N_C}{(4\pi)^2}\,,\quad
\al h=\frac{{u}\,N_F}{(4\pi)^2}\,,\quad
\al v=\frac{{v}\,N^2_F}{(4\pi)^2}\,.
\eeq
We have normalized the couplings with the appropriate powers of $N_C$ and $N_F$ preparing for the Veneziano limit to be considered below. 
Notice the additional power of $N_F$ in the definition of the scalar double-trace coupling,
meaning that $v/u$ becomes $\al v / (\al h\,N_F)$.
We also use  the shorthand notation $\beta_i\equiv\partial_t\alpha_i$ with $i=(g,y,h,v)$ to indicate the $\beta$-functions for the couplings \eq{couplings}. To obtain explicit expressions for these, we use the results \cite{Machacek:1983tz,Machacek:1983fi,Machacek:1984zw}. 

\subsection{Leading order}
We begin our reasoning with the RG flow for the gauge coupling to one-loop order using the $SU(N_C)$ Yang-Mills Lagrangean \eq{F2} coupled to $N_F$ fermions  \eq{F} in the fundamental representation,
\begin{equation}\label{dalpha1L}
\beta_g =\partial_t\, \alpha_g=-B\,\alpha_g^2\,.
\end{equation}
Note that a linear term $A\alpha_g$ is absent, unlike in \eq{dalpha}, as we strictly stick to four dimensions. To this order the gauge $\beta$-function \eq{dalpha1L} displays a doubly-degenerated fixed point at 
\beq 
\label{1}
\alpha_g^*=0
\eeq 
which is an UV fixed point for positive $B$. Provided that the coefficient $B$ is numerically very small, $|B|\ll 1$, however, we also have that $\partial_t\, \alpha_g\ll 1$ close to the Gaussian fixed point, indicating that the theory  might develop a non-trivial fixed point with 
\beq
\label{2}
0<\alpha_g^*\ll 1
\eeq
once higher loop terms are included. For $B>0$, \eq{1} corresponds to asymptotic freedom, in which case \eq{2} would then correspond to a conformal  infrared fixed point.\footnote{There is a vast body of work dealing with the availability of this Banks-Zaks type IR fixed point   \cite{Caswell:1974gg,Banks:1981nn} in the continuum and on the lattice (for recent overviews see \cite{Sannino:2009za,Kuti:2014epa} and references therein).} In turn, for $B<0$ asymptotic freedom is lost, the Gaussian fixed point becomes an IR fixed point,
and the theory may become asymptotically safe perturbatively at \eq{2}.
In our setup, the one-loop coefficient $B$ depends on both $N_C$ and $N_F$. Explicitly, $B=-\s043\eps$, where
\begin{equation}\label{eps}
\eps=\frac{N_F}{N_C}-\frac{11}{2}\,.
\end{equation}
For $\eps>0$, asymptotic freedom of the gauge sector is lost.  
The prerequisition for an asymptotically safe fixed point within the perturbative regime thus translates into 
\beq\label{small}
0\le\eps\ll 1\,.
\eeq 
Consequently, to achieve asymptotic safety in the gauge sector in a controlled perturbative manner, 
we must perform a Veneziano limit by sending both $N_C$ and  $N_F$ to infinity, but keeping their ratio $N_F/N_C$ fixed \cite{Veneziano:1979ec}. The parameter \eq{eps} thereby becomes continuous and can take any real value including \eq{small}. 
In most of the paper, the parameter $\eps$ will be our primary perturbative control parameter in the regime \eq{small}, except in Sec.~\ref{LargeCoupling} where we also discuss the regime where $\eps$
becomes large.

\subsection{Next-to-leading order}
We now must check whether this scenario can be realized upon the inclusion of higher loop corrections. At the next-to-leading (NLO) order in perturbation theory, which is two-loop, the RG flow for the gauge coupling takes the form
\begin{equation}\label{dalpha2}
 \partial_t\, \alpha_g=-B\,\alpha_g^2+C\,\alpha_g^3\,.
\end{equation}
As such, the gauge $\beta$-function \eq{dalpha2} may display three fixed points, a doubly-degenerated one at $\alpha_g^*=0$, and a non-trivial one at
\begin{equation}\label{FP2loop}
\alpha_g^*=B/C\,.
\end{equation}
The non-trivial fixed point is perturbative as long as $0\le\alpha_g^*\ll 1$ along with all the other possible couplings in the theory. In practice, this follows provided that $|B|\ll 1$ and $C$ of order unity, and $B/C>0$. For this fixed point to be an asymptotically safe one we must have $B<0$  and $C<0$.
However, in the absence of Yukawa interactions 
one finds  
$C=25$  \cite{Caswell:1974gg}  to leading order in $\eps$. Consequently, the would-be fixed point \eq{FP2loop} resides in the unphysical domain $\alpha^*_g<0$ where the theory is sick non-perturbatively, see e.g.~\cite{Dyson:1952tj}. 

\begin{figure*}[t]
\begin{center}
\includegraphics[width=.6\hsize]{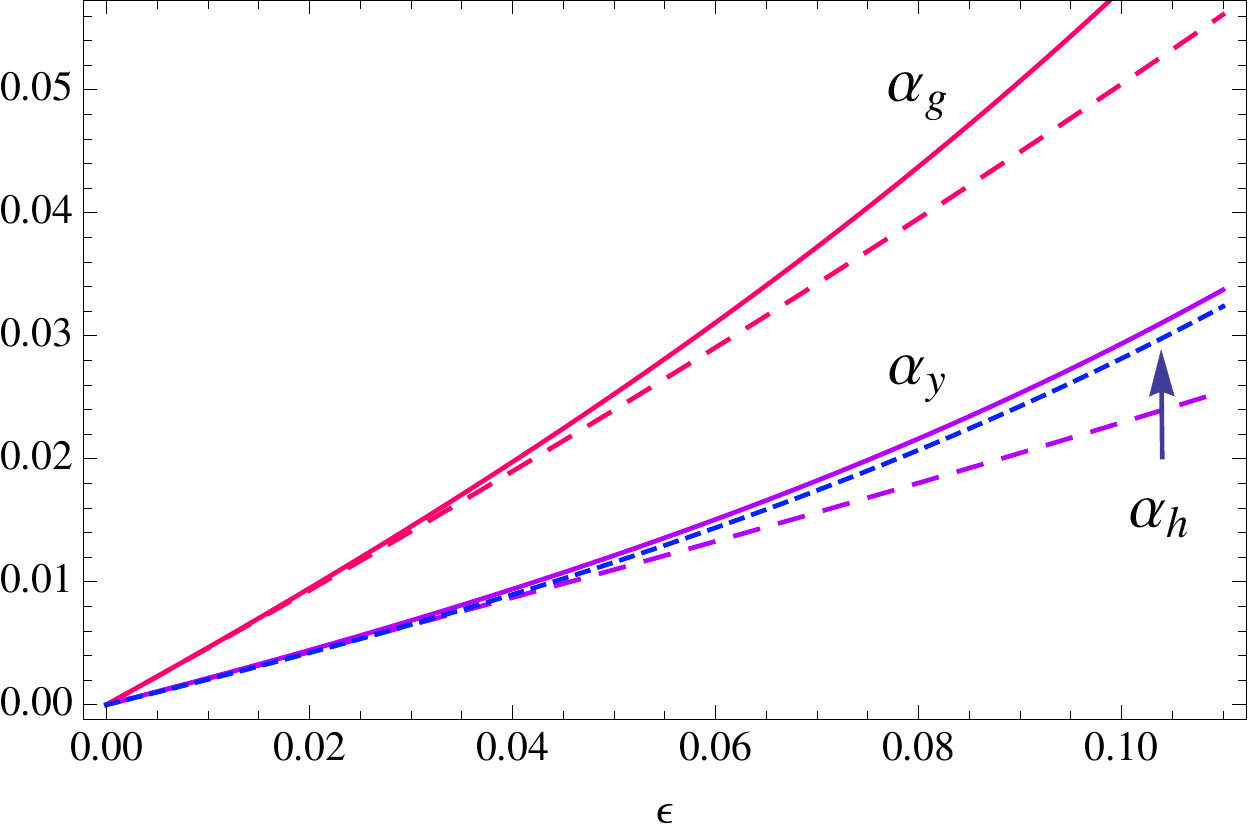}
\caption{\label{pAlphas} The coordinates of the UV fixed point as a function of $\eps$  at NLO (dashed) and NNLO (full and short-dashed lines). Gauge, Yukawa and scalar couplings are additionally shown in red, magenta and blue, respectively. NNLO corrections lead to a mild enhancement of the gauge and Yukawa couplings over their NLO values. The scalar and Yukawa fixed point couplings are nearly degenerate at NNLO.}
 \vskip-.3cm
\end{center}
\end{figure*} This conclusion changes as soon as Yukawa interactions are taken into consideration. The gauge $\beta$-function depends on the Yukawa coupling starting from the two-loop order.
For this reason, to progress, we must first evaluate the impact of non-trivial Yukawa couplings.
At the same time, since the fixed point for the gauge coupling depends on the Yukawa coupling, we must retain its RG flow to its first non-trivial order, which is one-loop.  Having introduced Yukawa couplings means that we also have dynamical scalars, \eq{Y}. The simplest choice here is to assume that the scalars are uncharged under the gauge group,  whence \eq{H}. Then neither the gauge nor the Yukawa RG flows depend on the scalar couplings at this order and we can neglect their contribution for now. This ordering 
of perturbation theory for the different couplings is also favored by considerations related to Weyl consistency conditions (see Sec.~\ref{Weyl}).
Hence, following this reasoning and  in the presence of the Yukawa term \eq{Y}, we end up with
 \beq\label{LOY}
\begin{array}{rcl}
\beta_g&=&
\displaystyle
\alpha_g^2 
\left\{
\frac{4}{3}\eps 
+ \left(25+\frac{26}{3}\eps\right) \alpha_g
-2
\left(\frac{11}{2}+\eps\right)^2 \alpha_y
\right\} \ ,
\\[2.5ex]
\beta_y&=&
\alpha_y\, \Big\{ (13 + 2 \eps) \,\alpha_y-6\,\alpha_g \Big\} \ .
\end{array}
\eeq
The coupled system \eq{LOY} admits three types of fixed points within perturbation theory \eq{small}. The system still displays a Gaussian fixed point
\beq\label{FP1}
(\alpha_g^*,\alpha_h^*)=(0,0)
\eeq
irrespective of the sign of $\eps$. For $\eps>0$,  neither the gauge coupling nor the Yukawa coupling are asymptotically free at this fixed point. Ultimately, this is related to the sign of either $\beta$-function being positive arbitrarily close to \eq{FP1}, making it an IR fixed point. A second, non-trivial fixed point is found as well,
$(\alpha_g^*,\alpha_h^*)\neq(0,0)$
which is an UV fixed point with coordinates
\beq\label{FP2}
\begin{array}{rcl}
\alpha_g^*&=&
\displaystyle
\frac{26\eps+ 4\eps^2}{57 - 46 \eps - 8 \eps^2}
=\frac{26}{57} \eps+ \frac{1424}{3249} \eps^2 + \frac{77360}{185193}\eps^3+{\cal O}(\eps^4)
\\[2ex]
\alpha_y^*&=& 
\displaystyle
\frac{12 \eps}{57 - 46 \eps - 8 \eps^2}=
\frac{4}{19}\eps + \frac{184}{1083}\eps^2 + \frac{10288}{61731} \eps^3 +{\cal O}(\eps^4)
\,.
\end{array}
\eeq
Numerically, the series \eq{FP2} reads
\beq
\begin{array}{rcl}
\alpha_g^*&=&
0.4561\, \eps+ 0.4383\,  \eps^2 + 0.4177\,  \eps^3+{\cal O}(\eps^4)\\[.5ex]
\alpha_y^*&=& 
0.2105\,  \eps + 0.1699\,  \eps^2 + 0.1667\,  \eps^3+{\cal O}(\eps^4)
\end{array}
\eeq
This UV fixed point is physically acceptable  in the sense that $(\alpha_g^*,\alpha_y^*)>(0,0)$ for $\eps>0$. It arises because the gauge and Yukawa couplings contribute with opposite signs to $\beta_g$ at the two-loop level, allowing for an asymptotically safe fixed point in the physical domain.  

As an aside, we also notice the existence of a second interacting fixed point  within perturbation theory located at 
\begin{equation}\label{BZ}
(\alpha_g^*,\alpha_y^*)=(-\frac{4 \eps}{75 + 26 \eps},0)\,.
\end{equation}
For $\eps>0$ this fixed point cannot be reached by any finite RG flow starting from the domain where $\alpha_g>0$ and  is unphysical. For $\eps<0$, \eq{BZ} takes the role of an interacting infrared fixed point  {\it \`a la} Caswell, Banks and Zaks 
\cite{Caswell:1974gg,Banks:1981nn}. Interacting IR fixed points play an important role in extensions of the standard model with a strongly interacting gauge sector and models with a composite Higgs (see~\cite{Sannino:2009za, Kuti:2014epa} and references therein).   The IR fixed point arises even in the absence of scalar fields, whereas the UV fixed point necessitates scalar matter with non-vanishing Yukawa interactions. We conclude that the IR fixed point \eq{BZ} is profoundly different from the UV fixed point \eq{FP2}.

  \begin{figure*}[t]
\begin{center}
\includegraphics[width=.6\hsize]{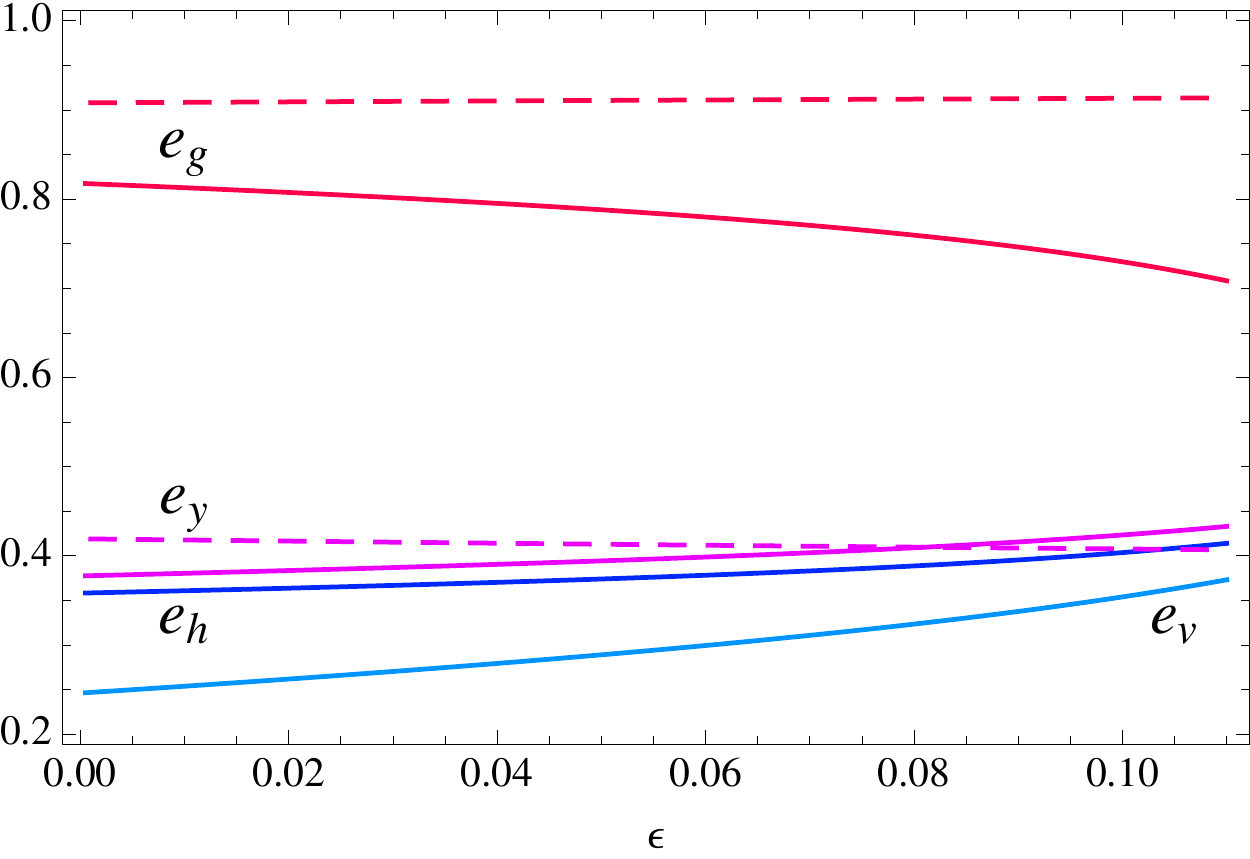}
\caption{\label{pLinear}Components of the eigenvectors \eq{normalisationNLO} and \eq{Eigendirection1}  (absolute values) corresponding to the relevant eigenvalue $\vartheta_1$ at the UV fixed point at NLO (dashed) and NNLO (full lines), respectively, as functions of $\eps$. From top to bottom,  the gauge, Yukawa and scalar, and double-trace scalar components are shown. Notice that the relevant eigendirection is largely dominated by the gauge coupling. Each component varies only mildly with $\eps$.}
 \vskip-.3cm
\end{center}
\end{figure*} 

Returning to our main line of reasoning, we linearize the RG flow in the vicinity of its UV fixed point \eq{FP2},
\begin{equation}\label{beta}
\beta_i=\sum_j M
_{ij}\,(\alpha_j-\alpha_j^*)+{\rm subleading}
\end{equation}
where $i=(g,y)$ and $M
_{ij}=\left.\partial\beta_i/\partial\alpha_j\right|_*$ denotes the stability matrix. The eigenvalues of $M$ are universal numbers and characterise the scaling of couplings in the vicinity of the fixed point. They can be found analytically.
The first few orders in \eq{eps} are
\beq\label{EV1}
\begin{array}{rcl}
\vartheta_1&=&
\displaystyle
-\0{104}{171}\eps^2 +\0{2296}{3249}\eps^3+\frac{1387768}{1666737}\eps^4+ {\cal{O}}(\eps^4)\\[2ex]
\vartheta_2&=&
\displaystyle
\ \ \, \0{52}{19}\eps +\0{9140}{1083}\eps^2+\0{2518432}{185193}\eps^3+ {\cal{O}}(\eps^4)\,.
\end{array}
\eeq
Numerically, the eigenvalues \eq{EV1} read
\beq
\begin{array}{rcl}
\vartheta_1&=&
\displaystyle
-0.608\,\eps^2 +0.707\,\eps^3-0.833\,\eps^4 +{\cal{O}}(\eps^5)\\[.5ex]
\vartheta_2&=&
\displaystyle
\ \ \,2.737\,\eps +  8.44\,\eps^2+13.599\,\eps^3+ {\cal{O}}(\eps^4)
\end{array}
\eeq
A few comments are in order. Firstly, the gauge-Yukawa system at NLO has developed a relevant and  an irrelevant eigendirection with eigenvalues $\vartheta_1<0$ and $\vartheta_2>0$, respectively.\footnote{Strictly speaking, there are two further marginal eigenvalues $\vartheta_{3,4}=0$ related to the scalar selfinteractions which we have taken as classical.}  

Secondly, the relevant eigenvalue $\vartheta_1$ is found to be of order $\eps^2$, whereas the irrelevant one is of order $\eps$ and thus parametrically larger. This is not a coincidence, and its origin can be understood as follows: All couplings  settle at values of order $\eps$ at the fixed point. Hence, $\beta_g\sim \eps^3$ and $\beta_y\sim \eps^2$ in the vicinity of the fixed point. 
The relevant eigenvalue originates primarily from the gauge sector, because asymptotic freedom is destabilised due to \eq{small}. Consequently, at the fixed point, the relevant eigenvalue must scale as $\vartheta_1\sim \partial\beta_g/\partial\eps|_* \sim\eps^2$. Conversely, the irrelevant eigenvalue scales as  $\vartheta_2\sim\partial\beta_y/\partial\eps|_* \sim\eps$. We conclude that the parametric dependence $\sim \eps^2$ of the relevant eigenvalue arises because the fixed point in the gauge sector stems from a cancellation at two-loop level. Conversely, the behaviour $\sim \eps$ of the irrelevant eigenvalue stems from cancellations at the one-loop level. This feature is a direct consequence of the vanishing mass dimension of the couplings. Asymptotic safety then follows as a pure quantum effect rather than through the cancelation of tree-level and one-loop terms.

Finally, introducing a basis in coupling parameter space as $\mathbf{a}=(\alpha_g,\alpha_y)^T$, we denote the relevant eigendirection at the UV fixed point  as 
\beq\label{normalisationNLO}
\mathbf{e_1}=(e_g,e_y)^T\,.
\eeq
The absolute values of its entries are shown in Fig.~\ref{pLinear} (dashed lines). We find that the relevant eigenvalue is largely dominated by the gauge coupling for all $\eps$. Furthermore, the relevant eigendirection is largely independent of $\eps$, up to $\eps<0.7$ where couplings become of order one. The domain of validity is further discussed in Sec.~\ref{validity} below.

\subsection{Next-to-next-to-leading order}\label{NNLO}
We  now move on to the next-to-next-to-leading order (NNLO) in the perturbative expansion where the scalar sector is no longer treated as exactly marginal.  Identifying a combined UV fixed point in all couplings becomes a  consistency check for asymptotic safety in the full theory.
 In practice, this amounts to adding the quartic selfinteraction terms \eq{U} and \eq{V}. The one-loop running of the quartic couplings given by
\bea\label{NLOH}
   \beta_h&=&
 -(11+ 2\eps) \,\alpha_y^2+4\alpha_h(\alpha_y+2\alpha_h)\,,\\
\label{NLOV}
  \beta_v&=&
12 \alpha_h^2  +4\al v \left(\alpha_v+ 4 \alpha_h+\alpha_y\right)\,.
\eea
For consistency, we also  include the two-loop corrections to the running of the Yukawa coupling  and the three-loop contributions to the running of the gauge coupling,  
\beq\label{NNLO}
 \begin{array}{rcl}
\Delta\beta_y^{(2)}&=&
\displaystyle
\alpha_y 
\left\{
\0{20 \eps-93}{6}\alpha_g^2 
+  (49 + 8 \eps) \alpha_g \alpha_y
-    \left(\0{385}{8} + \0{23}{2} \eps + \0{\eps^2}{2}\right) \alpha_y^2 
-  (44 + 8 \eps) \alpha_y \alpha_h 
+ 4 \alpha_h^2\right\} \\[2.5ex]
\Delta\beta_g^{(3)}&=&
\displaystyle
\alpha_g^2 \left\{ \left(\frac{701}{6}+  \0{53}{3} \eps - \0{112}{27} \eps^2\right) \alpha_g^2 - 
   \0{27}{8} (11 + 2 \eps)^2 \alpha_g \alpha_y + \0{1}{4} (11 + 2 \eps)^2 (20 + 3 \eps) \alpha_y^2\right\}  \,.
  \end{array}
\eeq
Several comments are in order. Firstly, both scalar $\beta$ functions are quadratic polynomials in the couplings and  display a Gaussian fixed point. The full system at NNLO then displays  a Gaussian fixed point 
$(\alpha_g^*,\alpha_y^*,\alpha_h^*,\alpha_v^*)=(0,0,0,0)$ as it must. 

Secondly, and unlike all other $\beta$-functions, the RG flow \eq{NLOV} for the double-trace coupling shows no explicit dependence on $\eps$.
Also, to leading order in $1/N_C$ and  $1/N_F$, \eq{NLOV}  stays quadratic in its coupling to all loop orders \cite{Pomoni:2008de}. 

Finally, and most notably, the $\beta$-functions of the gauge, Yukawa and single-trace scalar coupling remain independent of the double-trace scalar coupling $\al v$. 
In consequence 
the dynamics of $\al v$ largely decouples from the system and it acts like a spectator without influencing the build-up of the asymptotically safe UV fixed point in the gauge-Yukawa-scalar subsector. Its own RG evolution is primarily fueled by the Yukawa and the single trace coupling, and as such indirectly sensitive to the gauge-Yukawa fixed point. In turn, the scalar coupling $\al h$ couples back into the Yukawa coupling, though not into the gauge coupling.

Since the RG flows at NNLO partly factorize, we can start by first considering the corrections to \eq{FP2} induced by the scalar coupling $\al h$. This leads to UV fixed points in the gauge-Yukawa-scalar subsystem
\beq
(\alpha_g^*,\alpha_y^*,\alpha_h^*)\neq(0,0,0)\,.
\eeq
In the background of the gauge-Yukawa fixed point \eq{FP2}, the RG flow \eq{NLOH} for $\al h$ admits two fixed points $\al {h2}^*<0<\al {h1}^*$, with 
\beq\label{h1h2}
\al {h1,h2}^*=(\pm\sqrt{23}-1)\,\frac{\eps}{19}+{\cal O}(\eps^2)\,,
\eeq
irrespective of $\al v$. Inserting $\al {h1}^*$ together with \eq{FP2} into \eq{NLOV} we then also find two solutions for the double trace coupling, with $\al {v2} ^*<\al {v1} ^*$. These are discussed in more detail below. Conversely, inserting $\al {h2}^*$ together with \eq{FP2} into \eq{NLOV} does not offer a fixed point for $\al v$. We  conclude that the fixed point $\al h^*\equiv \al {h1}^*>0$ is the sole value for the coupling $\al h$ which leads to a UV fixed point in the scalar subsystem. We return to this in Sec.~\ref{discuss}.

Using \eq{LOY}, \eq{NLOH}, and \eq{NNLO}, the coordinates of the gauge, Yukawa and scalar coupling 
are obtained analytically and can be expressed as a power series in $\eps$ starting as
\beq\label{alphaNNLO}
\begin{array}{rcl}
\alpha_g^*&=&
\displaystyle
\frac{26}{57}\,\eps+ \frac{23 (75245 - 13068 \sqrt{23})}{370386}\,\eps^2
+{\cal O}(\eps^3)
\\[1.ex]
\alpha_y^*&=&
\displaystyle
\frac{4}{19}\,\eps+\left(\frac{43549}{20577} - \frac{2300 \sqrt{23}}{6859}\right)\,\eps^2
+{\cal O}(\eps^3)
\\[1ex]
\alpha_h^*&=&
\displaystyle
\frac{\sqrt{23}-1}{19}\,\eps+\frac{1168991 - 202249 \sqrt{23}}{82308 \sqrt{23}}\,\eps^2
+{\cal O}(\eps^3)\,.
\end{array}
\eeq
Numerically, the first few orders in the $\eps$-expansion read
\beq\label{NNLOseries}
\begin{array}{rcl}
\alpha_g^*&=&\; \; \,0.4561\,\eps+0.7808 \,\eps^2+
3.112\,\eps^3+{\cal O}(\eps^4)\\[.5ex]
\alpha_y^*&=&\; \; \,0.2105\,\eps+0.5082\,\eps^2+
2.100\,\eps^3+{\cal O}(\eps^4)\\[.5ex]
\alpha_h^*&=&\,  \; \;0.1998\,\eps+0.5042\,\eps^2+
2.045\,\eps^3+{\cal O}(\eps^4)\,.
\end{array}
\eeq
The addition of the scalar selfcouplings has led to a physical fixed point $\alpha_h^*>0$ of order $\eps$. NNLO corrections to $\alpha_g^*$ and $\alpha_y^*$ arise only starting at order $\eps^2$ without altering the NLO fixed point \eq{FP2}. 
Performing the expansion \eq{NNLOseries} to high orders in $\eps$ one  finds  its radius of convergence as
\beq
\label{max}
\eps\le \eps_{\rm max}= 0.117\cdots
\eeq
At $\eps_{\rm max}$ in \eq{max}, the NNLO equations display a bifurcation and the UV fixed point ceases to exist through a merger with a non-perturbative IR fixed point, and the relevant eigenvalue disappears at $\eps_{\rm max}$. The merger at $\eps_{\rm max}$ indicates that our working assumption \eq{small} should be superseeded by $0<\eps\ll \eps_{\rm max}$.

Since $\al v$ does not contribute to the RG flow of the subsystem $(\alpha_g,\alpha_y,\alpha_h)$, the computation of scaling exponents equally factorizes.  
Linearizing the RG flow in the vicinity of the fixed point, we  have \eq{beta}
where now $i=(g,y,h)$. 
The  eigenvalues $\vartheta_n$ are found analytically as a power series in $\eps$, the first few orders of which are given by
\beq\label{thetaNNLO}
\begin{array}{rcl}
\vartheta_1&=&
\displaystyle
-\frac{104}{171}\,\eps^2+\frac{2296}{3249}\,\eps^3+\frac{4531558295989 - 922557832416 \sqrt{23}}{46931980446}\,\eps^4+{\cal O}(\eps^5)\\[2ex]
\vartheta_2&=&
\displaystyle
\ \ \,\frac{52}{19}\, \eps +\frac{136601719 - 22783308 \sqrt{23}}{4094823}\,\eps^2+{\cal O}(\eps^3)\\[2ex]
\vartheta_3&=&
\displaystyle
\ \ \frac{16\sqrt{23}}{19}\,\eps+4\frac{217933589 \sqrt{23}-695493228}{94180929}\,\eps^2
+{\cal O}(\eps^3)
\end{array}
\eeq
For the relevant eigenvalue, we notice that the first two non-trivial orders have remained unchanged. Numerically, the eigenvalues  read
\beq
\begin{array}{rcl}
\vartheta_1&=&-0.608\,\eps^2+0.707 \,\eps^3+2.283\,\eps^4+\cdots\\
\vartheta_2&=&\ \ \,2.737\, \eps\  +6.676\,\eps^2+\cdots\\
\vartheta_3&=&\ \ \,4.039\,\eps\ +14.851\,\eps^2+\cdots\,.
\end{array}
\eeq
The cubic and quartic corrections to the relevant eigenvalue both arise with a sign opposite to the leading term, which is responsible for the smallness of $\vartheta_1$ even for moderate values of $\eps$. In turn, the irrelevant eigenvalues receive larger corrections and reach values of the order of $0.1\div 1$ for moderate $\eps$.

   \begin{figure*}[t]
\begin{center}
\includegraphics[width=.6\hsize]{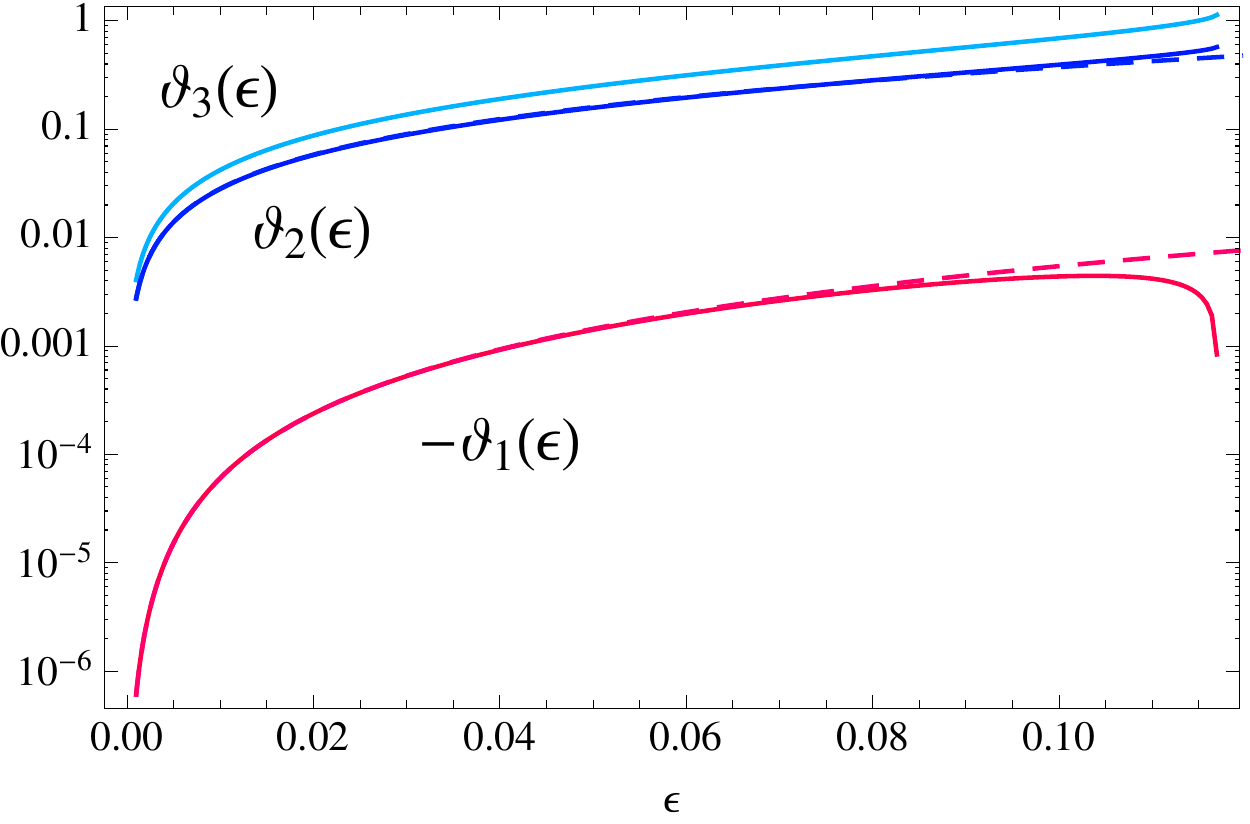}
\caption{\label{pTheta}  Universal  eigenvalues $\vartheta_1<0<\vartheta_2<\vartheta_3$ at the UV fixed point at NLO (dashed lines) and NNLO (full lines) as function of $\eps$ \eq{eps}. For all technical purposes, the eigenvalues are dominated by the NLO values except for the close vicinity of $\eps\approx\eps_{\rm max}$.}
 \vskip-.3cm
\label{pScaling}
\end{center}
\end{figure*}

We now discuss the role of the double-trace scalar coupling $\al v$. Its fixed points are entirely induced by the fixed point in the gauge-Yukawa-scalar subsystem  \eq{alphaNNLO}. 
Two solutions are found,
\bea\label{v1v2}
\label{v1}
\al {v1,v2} ^*&=&
-\frac{1}{19} \left(2 \sqrt{23}\mp\sqrt{20 + 6 \sqrt{23}}\right)\,\eps
+{\cal O}(\eps^2)
\eea
Numerically  $\al {v1} ^*=-0.1373\,\eps$ and $\al {v2} ^*=-0.8723\,\eps$, up to quadratic corrections in $\eps$. In principle, either of these fixed points can be used in conjunction with \eq{alphaNNLO} to define the combined UV fixed point of the theory.
We also note that 
\begin{equation}
\al {h1}^*+\al {v2} ^*<0<\al {h1}^*+\al {v1} ^*\,,
\end{equation}
showing that the scalar field potential is  bounded (unbounded) from below for the latter (former) choice of couplings.\footnote{An inspection of radiative corrections confirms that the effective potential for the scalar fields 
is stable classically and quantum-mechanically  \cite{LMS:2014}.}
At the fixed point $\al {v1} ^*$, the invariant \eq{V} becomes perturbatively irrelevant and adds a positive scaling exponent to the spectrum. Conversely, at the fixed point  $\al {v2} ^*$,  the invariant \eq{V} has become perturbatively relevant. Since the RG flow \eq{NLOV} is quadratic in the coupling $\al v$ to all orders, the corresponding scaling exponents are equal in magnitude with opposite signs,
       \beq
\vartheta_4= \frac{8\,\eps}{19} \sqrt{20 + 6 \sqrt{23}}+{\cal O}(\eps^2)\,.
\eeq
Numerically, $\vartheta_4=2.941\,\eps+{\cal O}(\eps^2)$. The occurence of an additional negative eigenvalue is  
induced 
by the interacting UV fixed point \eq{alphaNNLO}.

We may introduce a basis in coupling parameter space as $\mathbf{a}=(\alpha_g,\alpha_y,\al h,\al v)^T$ to denote the eigendirections at the UV fixed point  as $\mathbf{e_i}$. They obey the eigenvalue equation $M \mathbf{e_i}=\vartheta_i \,\mathbf{e_i}$. 
The normalised relevant eigendirection $\mathbf{e_1}$ corresponding to the UV attractive eigenvalue  has the components 
\begin{equation}\label{Eigendirection1}
\mathbf{e_1}=(e_g,e_y,e_h,e_v)^T
\end{equation}
whose values as function of \eq{eps} are shown in Fig.~\ref{pLinear}. Clearly, even at NNLO the gauge coupling dominates the relevant eigendirection. We also find that the eigendirection corresponding to $\vartheta_4$ is independent of the gauge, Yukawa and scalar coupling,  $\mathbf{e_4}=(0,0,0,1)$.

\subsection{UV scaling and Landau pole}\label{discuss}
We  summarize the main picture (see also Tab.~\ref{Tab1}). The asymptotically safe UV fixed point \eq{FP2} in the gauge-Yukawa system bifurcates into several fixed points once the scalar fluctuations are taken into account. In addition to the universal eigenvalues of the gauge-Yukawa system $\vartheta_1$ and $\vartheta_2$, the scalar sector add the eigenvalues $\pm \vartheta_3$ and $\pm\vartheta_4$. To the leading non-trivial order in $\eps$, these are
\beq\label{thetavalues}
\begin{array}{rcl}
\vartheta_1&=&-0.608\,\eps^2+{\cal O}(\eps^3)
\\
\vartheta_2&=&\ \ \,2.737\, \eps+{\cal O}(\eps^2)
\\
\vartheta_3&=&\ \ \,4.039\,\eps+{\cal O}(\eps^2)
\\
\vartheta_4&=&\ \ \,2.941\,\eps+{\cal O}(\eps^2)\,.
\end{array}
\eeq
Complete asymptotic safety, e.g.~asymptotically safe UV fixed points in all couplings, is achieved at two UV fixed points, FP${}_1$ and FP${}_2$.
At FP${}_1$, both scalar invariants are perturbatively irrelevant and the eigenvalue spectrum is \beq\label{evFP1}
\vartheta_1<0<\vartheta_2<\vartheta_4<\vartheta_3\,.
\eeq
At FP${}_2$, the partly decoupled double-trace scalar interaction term with coupling $\al v$ becomes relevant in its own right, and the eigenvalue spectrum, instead, reads 
\beq\label{evFP2}
-\vartheta_4<\vartheta_1<0<\vartheta_2<\vartheta_3\,. 
\eeq
At either of these, the UV limit can be taken. We recall that the scalar field potential is stable (unstable) at FP${}_1$ (FP${}_2$), indicating that FP${}_1$ corresponds to a physically acceptable theory at highest energies.
Finally, a fixed point FP${}_3$ exists for the gauge-Yukawa-scalar subsystem. Here, the scalar coupling $\al h$ becomes relevant and the eigenvalue spectrum reads \beq\label{evFP3}
-\vartheta_3<\vartheta_1<0<\vartheta_2\,.
\eeq
Here, however, asymptotic safety is not complete. In fact, the $\beta$-function for the double-trace scalar coupling does not show a fixed point in perturbation theory as it remains strictly positive, leading to Landau poles $\al v\to \pm\infty$  in the IR and in the UV.
 In the UV, this regime resembles the $U(1)$ or Higgs sector of the standard model. In either case it is no longer under perturbative control. Unless strong-coupling effects resolve this singularity in the UV, this behaviour  implies a  limit of  maximal UV extension of the model close to FP${}_3$. 

\begin{center}
\begin{table}
\begin{tabular}{c|cccc|cc}
 \hline\hline
\multirow{2}{*}{{}\ \ fixed point\ \ }
&
\multicolumn{4}{c}{{}\ \ \ couplings \ \ \ }
&\multicolumn{2}{c}{{}\ \ \ eigenvalues \ \ \ }
 \\ 
 \cline{2-7}
&\ \ gauge\ \ &Yukawa&\ \ scalar\ \ &double-trace
&{}\ \ \  \ \ relevant{}\ \ \ \ \  
&{}\ \ \ \ \ irrelevant{}\ \ \ \ \ \\ \hline
FP${}_1$     
&$\ \  \alpha_{g}^*\ \  $
&$\ \  \alpha_{y}^*\ \  $
&$\ \  \alpha_{h1}^*\  \ $
&$\ \  \alpha_{v1}^*\  \ $
&$\vartheta_1$ 
& $\vartheta_2,\ \vartheta_3,\ \vartheta_4$ \\ 
 FP${}_2$     
&$\alpha_{g}^*$
&$\alpha_{y}^*$
&$\alpha_{h1}^*$
&$\alpha_{v2}^*$
& $-\vartheta_4,\ \vartheta_1$ &$\vartheta_2,\ \vartheta_3$ \\ 
 FP${}_3$     
&$\alpha_{g}^*$
&$\alpha_{y}^*$
&$\alpha_{h2}^*$&\ \ \ Landau pole\ \ \ &
 $-\vartheta_3,\ \vartheta_1$ &$\vartheta_2$\\  \hline\hline
\end{tabular}
\caption{\label{Tab1}
 Summary of UV fixed points of the gauge-Yukawa theory and the number of (ir-)relevant eigenvalues. To leading non-trivial order, the fixed point values are given by \eq{FP2}, \eq{h1h2}, \eq{v1v2} and the exponents by \eq{thetavalues}.}
\end{table} 
\end{center}

\subsection{UV critical surface}

The existence of relevant and irrelevant direction in the UV implies that the short-distance behaviour of the theory is described by a lower-dimensional UV critical surface. We discuss the NLO case and FP${}_1$ at NNLO in detail. This is straightforwardly generalised to the case with two relevant eigendirections.  On the critical surface in coupling constant space, we may express the RG running of the irrelevant  coupling, say $\alpha_i$ with $(i=y,h,v)$, 
in terms of the relevat one, say $\alpha_g$, leading to relations 
\beq\label{hyper}
\begin{array}{rcl}
\alpha_i&=&F_i(\alpha_g)\,.
\end{array}
\eeq
To see this more explicitly, 
we integrate the RG flow in the vicinity of the UV fixed point 
to find
the general solution 
\begin{equation}\label{lambda}
\alpha_i(\mu)=\alpha_i^*+\sum_n \,c_n\, V^n_i\, \left(\frac{\mu}{\Lambda_c}\right)^{\vartheta_n}+{\rm subleading}\,.
\end{equation}
Here, $\Lambda_c$ is a reference energy scale, $\vartheta_n$ are the  eigenvalues of the stability matrix $
M$, $V^n$ the corresponding eigenvectors, and $c_n$ free parameters. The eigenvectors generically mix all couplings.  At NLO the eigenvalues obey $\vartheta_1<0<\vartheta_2$. For all coupling $\alpha_i$ to reach the UV fixed point with increasing RG scale $1/\mu\to 0$ we therefore must set the free parameter $c_2=0$. Conversely, the parameter $c_1$ remains undetermined and should be viewed as a free parameter of the theory. Since this holds true for each coupling $\alpha_i$, we can eliminate $c_1$ from  \eq{lambda} to express the irrelevant coupling in terms of the relevant one. At NLO, one finds
\beq\label{hypercritical1}
F_y(\alpha_g)=\left(\frac{6}{13}-\frac{88\eps}{507}
\right)\alpha_g+\frac{8}{171} \eps^2+{\cal O}(\eps^3)
\eeq
to the first few   orders in an expansion in $\eps$. At NNLO, the hypercritical surface is extended and receives corrections due to the scalar couplings. At FP${}_1$, for example, one finds
\beq\label{hyper3}
\begin{array}{rcl}
F_y(\alpha_g)&=&\ \ \, (0.4615  + 0.6168 \,\eps)\,\al g -0.1335\,\eps^2+{\cal O}(\eps^3)\\[1ex]
F_h(\alpha_g)&=&\ \ \, (0.4380 +    0.5675\, \eps)\,\al g -0.09658\,  \eps^2+{\cal O}(\eps^3)\,,\\[1ex]
F_v(\alpha_g)&=&-(0.3009 + 0.3241\,\eps)\,\al g + 0.1373 \,\eps + 0.3828\, \eps^2+{\cal O}(\eps^3)
\end{array}
\eeq
to the first few   orders in $\eps$, and in agreement with \eq{hypercritical1} to order $\eps$. The significance of the UV critical surface  is that couplings can reach the UV fixed point only along the relevant direction, dictated by the eigenperturbations. This imposes a relation between the relevant and the irrelevant coupling, both of which scale out of the fixed point with the same scaling exponent $\vartheta_1$. On the critical surface and close to the fixed point, the gauge coupling evolves as
\begin{equation}\label{lambdaHS}
\alpha_g(\mu)=\alpha_g^*+\Big(\alpha_g(\Lambda_c)-\alpha_g^*\Big)\, \left(\frac{\mu}{\Lambda_c}\right)^{\vartheta_1(\eps)}
\end{equation}
and the irrelevant couplings follow suit 
via \eq{hyper} with \eq{hypercritical1} and \eq{hyper3},
and with $\alpha_g^*$ and $\vartheta_1$ given by the corresponding expressions at NLO and NNLO, respectively.

   \begin{figure*}[t]
\begin{center}
\includegraphics[width=.6\hsize]{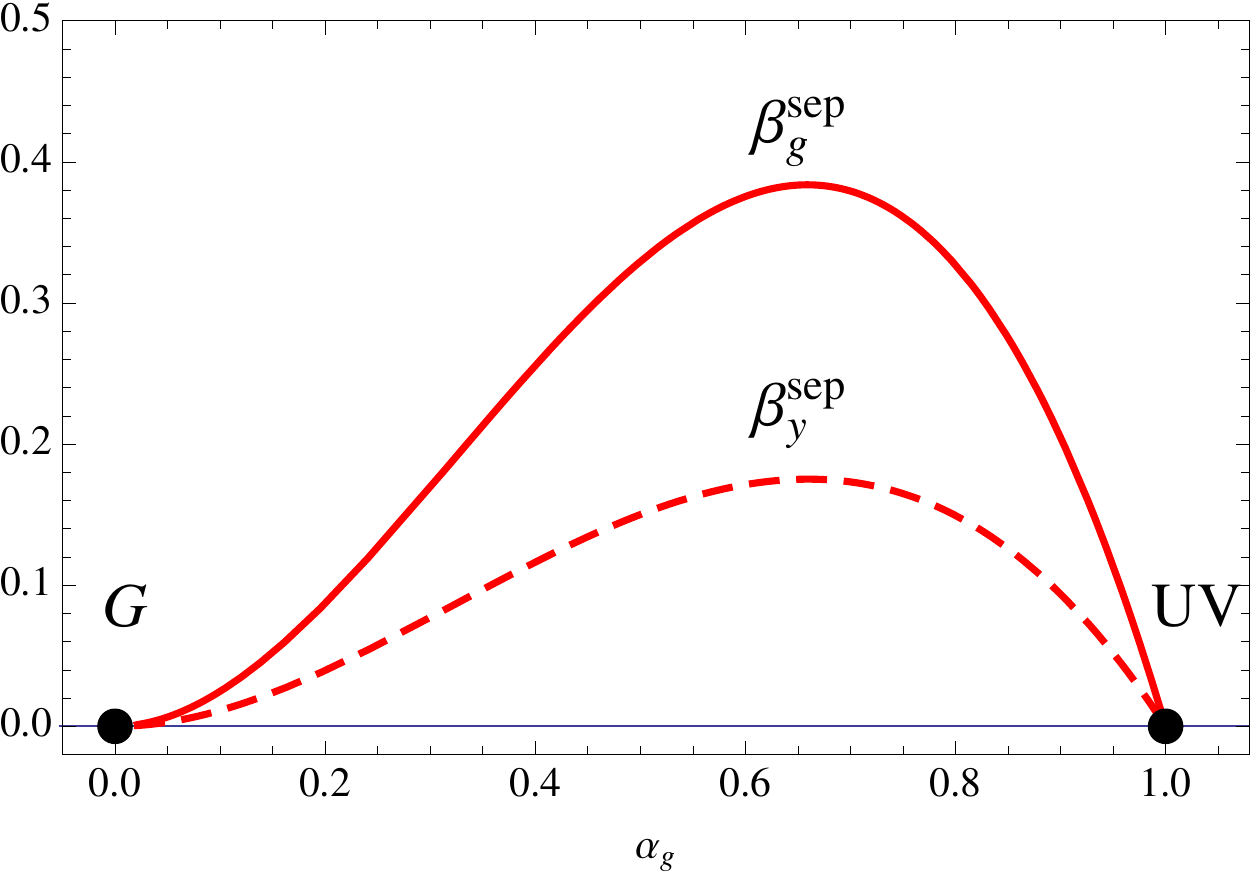}
\caption{\label{pBetaSep}  The gauge and Yukawa $\beta$-functions  projected along the separatrix of the phase diagram given in Fig.~\ref{pPD} (NLO with $\eps = 0.05$), also showing the Gaussian and the UV fixed point. For better display, we have rescaled $\beta\to\beta/(\alpha_g^*)^3$, and $\alpha_g\to \alpha_g/\alpha_g^*$; see main text.}
 \vskip-.5cm
\end{center}
\end{figure*} 

\subsection{Phase diagram}\label{PD}

In Fig.~\ref{pPhase}, we show the phase diagram of the asymptotically safe gauge-Yukawa theory for small couplings at NLO, where we have 
set $\eps=0.05$. 
The RG trajectories are obtained from integrating \eq{LOY} with arrows pointing towards the IR. For small couplings, one observes the Gaussian and the UV fixed points. At vanishing Yukawa (gauge) coupling, the RG equations \eq{LOY} become infrared free in the gauge (Yukawa) coupling, corresponding to the thick red horizontal (vertical) trajectory in Fig.~\ref{pPhase}. This makes the Gaussian fixed point IR attractive in both couplings. The UV fixed point has a relevant and an irrelevant direction, corresponding to the two thick red trajectories one of which is flowing out of and the other into the UV fixed point. These trajectories are distinguished in that they also divide the phase diagram into four regions $A, B, C$ and $D$. 

The  trajectory connecting the UV with the Gaussian fixed point is a separatrix, which defines the boundary between the regions $A$ and $C$, and $B$ and $D$. Close to the UV fixed point, its coordinates are given by   the UV critical surface \eq{hypercritical1}. 
The RG flow along the separatrix is given analytically to very good accuracy by \eq{lambdaHS} and \eq{hypercritical1}, and for suitable initial conditions $\alpha_g(\mu=\Lambda_c)$ on the separatrix. Note that due to the smallness of the relevant UV eigenvalue compared to the irrelevant one, $|\vartheta_1/\vartheta_2|=\frac29 \eps+{\cal O}(\eps^2)$, the RG flow runs very slowly along the separatrix. In turn, trajectories entering into the separatrix (with decreasing RG scale $\mu$) run much faster, reflected by their near-perpendicular angle between these trajectories and the separatrix; see  Fig.~\ref{pPhase}. 
  
Trajectories in region $A$ run towards the Gaussian FP in the IR, and towards strong Yukawa coupling in the UV. Trajectories in $B$ run towards a strong coupling regime in the IR. In region $C$, trajectories approach the Gaussian FP in the IR limit. Finally, in region $D$ trajectories approach a strongly coupled regime in the IR, outside the domain of applicability of our equations. Notice that the Gaussian fixed point is attractive in all directions.  Hence, asymptotically safe trajectories emanating from the UV fixed point  either run towards a weakly coupled phase controlled by the Gaussian fixed point in the deep IR, or towards a strongly coupled QCD-like phase characterised by chiral symmetry breaking and confinement.
More generally, in Fig.~\ref{pPhase}, the  boundary between weakly ($A$ and $C$) and strongly ($B$ and $D$) coupled phases at low energies is given by the UV irrelevant direction, i.e.~the two full (red) trajectories running into the UV fixed point.

In Fig.~\ref{pBetaSep} , we show the gauge and Yukawa $\beta$-functions projected along the separatrix as functions of the gauge coupling, 
\beq
\begin{array}{rcl}
\beta^{\rm sep}_g(\al g)&\equiv&\beta_g(\al g,\al y=F_y(\al g))\\[1.5ex]
\beta_y^{\rm sep}(\al g)&\equiv&\beta_y(\al g,\al y=F_y(\al g))
\end{array}
\eeq
also using \eq{hyper} with \eq{hypercritical1}. Both of them display the Gaussian and the UV fixed point. We also recover the UV relevant eigenvalue 
\beq
\vartheta_1=\frac{d\beta^{\rm sep}_i}{d\al g}\Big|_*=
\frac{\partial\beta_i}{\partial \al g}\Big|_*
+\frac{\partial F_y}{\partial \al g}\frac{\partial\beta_i}{\partial \al y}\Big|_*
\eeq
from either of these $ (i=g,y)$. Close to the UV fixed point  the RG running is power-like. Close to the IR fixed point, the running becomes logarithmic. Quantitatively, along the separatrix the crossover from UV scaling to IR scaling takes place once couplings are reduced to about $\sim 65\%$ of their UV fixed point values.

\subsection{Mass terms and anomalous dimensions}
 If mass terms are present, their multiplicative renormalisation is induced through the RG flow of the gauge, Yukawa, and scalar couplings. We now discuss the scaling associated to the scalar wave function renormalization, the scalar mass, and the addition of the fermion mass operator. The former is identified with the  anomalous dimension $\gamma_H$ for the scalar wave function renormalization\footnote{The bare and renormalized fields here are related via $H_B = Z_H^{\frac{1}{2}} H$. Also the wave function definition in \cite{Machacek:1983fi} is the inverse of the one here.} $Z_H$,
\begin{equation}
\label{adimensionX}
\Delta_H=1+\gamma_H \hspace{1mm}, \qquad \gamma_H\equiv - \frac{1}{2}\frac{d\ln Z_H}{d\ln\mu}\,.
\end{equation}
Within perturbation theory, the one and two loop contributions to \eq{adimensionX} read (see e.g.~\cite{Luo:2002ti})
\begin{eqnarray}
\label{gammaH1}\gamma_H^{(1)}&=&\alpha_y\ ,\\
\label{gammaH2}\gamma_H^{(2)}&=& 
-\032\left(\0{11}{2}+\eps\right)\alpha_y^2
+\052\alpha_y\alpha_g
+2\alpha_h^2 \,.
\end{eqnarray}
Inserting the UV fixed point  FP${}_1$ and expanding the anomalous dimension in powers of $\eps$, we find
\beq
\gamma_H=\frac{4 \eps}{19} +\frac{14567 - 2376 \sqrt{23}}{6859}\,\eps^2+{\cal O}(\eps^3)\,.
\eeq
Notice that the leading and subleading terms are both 
positive. 
 The anomalous dimension for the scalar mass term $m$ can be derived from the composite operator $\sim m^2\,\Tr\, H^\dagger H$. Introducing $\gamma_m=\frac 12 d\ln m^2/d \ln \mu$ 
one finds 
the mass anomalous dimension 
\beq\label{gammam}
\gamma^{(1)}_m=2\al y+4\al h+ 2\al v
\eeq
to one-loop order. We find that \eq{gammam} becomes as large as $\gamma^{(1)}_m\approx 0.09$ for $\eps\approx 0.1$ at FP${}_1$. Evidently, the loop corrections remain small compared to the tree level term leaving $m^2=0$ as the sole fixed point within perturbation theory. Analogously, the anomalous dimension for the fermion mass operator is defined as
\begin{equation}
\label{adimensionF}
\Delta_F=3 - \gamma_F \hspace{1mm}, \qquad \gamma_F\equiv  \frac{d\ln M}{d\ln\mu} 
\end{equation}
where $M$ stands for the fermion mass. Within perturbation theory, the one and two loop contributions read
\begin{eqnarray}
\label{gammaF1}\gamma_F^{(1)}&=&3 \alpha_g-\alpha_y\, \left(\0{11}{2}+\eps\right)\ ,\\
\label{gammaF2}\gamma_F^{(2)}&=&\left(44+8\eps\right)\alpha_g\alpha_y
+\left(\0{31}{4}-\053\eps\right)\alpha_g^2
+\014\left (\0{11}{2} + \eps\right) \left(\0{23}{2} + \eps\right)\alpha_y^2\,.
\end{eqnarray}
Inserting the UV fixed point  FP${}_1$ and expanding in $\eps$ we find
\beq
\gamma_F=\frac{4 \eps}{19} +\frac{4048 \sqrt{23}-59711}{6859}\,\eps^2+{\cal O}(\eps^3)\,.
\eeq
The leading and subleading terms are both positive.
Interestingly, to one-loop order, the scalar anomalous dimension and the fermion mass anomalous dimension coincide in magnitude. The quantum corrections are bounded, $|\gamma_H^{(1)} |<1/40$. 
We stress that the leading order results are entirely fixed by the NLO fixed point \eq{FP2}, and insensitive to the details of the scalar sector. The latter only enter starting at order $\eps^2$.

\section{Consistency}\label{consistency}
In this section, we discuss aspects of consistency and the validity of results.

 \subsection{Stability} In the LO, NLO and NNLO approximations, we have retained the $\beta$-functions of the gauge, Yukawa and quartic couplings at different loop levels within perturbation theory. As we have argued, the ordering as shown in Tab.~\ref{Tab} is dictated by the underlying dynamics towards asymptotic safety, centrally controlled by the gauge coupling. 
 
 The selfconsistency of our reasoning is confirmed a posteriori by the stability of the result. 
 Firstly, 
the leading coefficients in $\epsilon$ of the NLO fixed point $\alpha^*_g$ and $\alpha^*_y$ remain numerically unchanged at NNLO, see \eq{FP2} and \eq{alphaNNLO}. We therefore expect that all coefficients up to  $\eps^2$ of $\alpha^*_g$ and $\alpha^*_y$ and the $\eps$ coefficient of $\alpha^*_h$ and $\alpha^*_v$ in \eq{alphaNNLO}, \eq{h1h2} and \eq{v1v2} remain unchanged beyond NNLO. 
Secondly, the stability also extends to the universal eigenvalues. Interestingly, here, the first two non-trivial coefficients (up to order $\eps^3$)  for the relevant eigenvalue $\vartheta_1$ at NLO agree with the NNLO coefficients, see \eq{EV1} and \eq{thetaNNLO}. For the leading irrelevant eigenvalue $\vartheta_2$, this agreement holds for the leading (order $\eps$) coefficient. 

All couplings of the theory have become fully dynamical at NNLO. At N${}^3$LO in the expansion, no new consistency conditions arise. Instead, higher loop corrections will lead to higher order corrections in the results established thus far. Based the observations above, we expect that all coefficients up to $\eps^4$ $(\eps^2)$ $[\eps]$ of the universal eigenvalues $\vartheta_1$ $(\vartheta_2)$ $[\vartheta_{3,4}]$ at NNLO in \eq{thetaNNLO} are unaffected at N${}^3$LO and beyond.

\subsection{Weyl consistency} \label{Weyl}
At a more fundamental level an argument known as Weyl consistency conditions \cite{Jack:1990eb,Jack:2013sha} lends a formal derivation of this hierarchical procedure of Tab.~\ref{Tab}. 
Replacing the couplings \eq{couplings} by the set $\{ g_i \} \equiv \{ g, y,u,v\}$ with $\beta$ functions $\beta_i=dg_i/d\ln\mu$,
the Weyl consistency condition 
\begin{equation}
	\frac{\partial \beta^j}{\partial g_i} = \frac{\partial \beta^i}{\partial g_j}
	\label{eq:integrabilitycondition}
\end{equation}
relates partial derivatives of the various $\beta$ functions to each other, and $\beta^i \equiv \chi^{ij} \beta_j$. The functions $\chi^{ij}$ plays the role of a metric in the space of couplings. The relations are expected to hold in the full theory, and hence it is desirable to obey \eq{eq:integrabilitycondition} even within finite approximations. The crucial point here is that the metric itself is a function of the couplings. Therefore, a consistent solution  to \eq{eq:integrabilitycondition} will generically relate different orders within a {\it na\"ive} fixed-order perturbation theory. In \cite{Antipin:2013sga} it was  shown that these conditions hold for the standard model. For the gauge-Yukawa theory studied here, the metric $\chi$ has been given explicitly in \cite{Antipin:2013pya} showing that the ordering  laid out in Tab.~\ref{Tab} is consistent with  \eq{eq:integrabilitycondition}.

\begin{center}
\begin{table}
\begin{tabular}{c|ccc}
 \hline\hline
 {}\ \ \ coupling\ \ \ &\multicolumn{3}{c}{{}\ \ \ order in perturbation theory\ \ \ }\\ \hline
 $\alpha_g$&1&2&3 \\
  $\alpha_y$ &0&1&2 \\
   $\alpha_h$ &0&0&1\\
    $\alpha_v$ &0&0&1 \\ \hline
    \ \ approximation level\ \ &{}\ \ \ \ \ LO\ \ \ \ \  &\ \ NLO\ \  &NNLO\\ \hline\hline
\end{tabular}
\caption{\label{Tab} Relation between approximation level and the loop order to which couplings are retained in perturbation theory.}
\end{table} 
\end{center}

 \subsection{Universality}

For our explicit computations we have used known RG equations in the MS-bar regularisation scheme.  In general, the expansion coefficients of $\beta$-functions are non-universal numbers and depend on the adopted scheme. On the other hand, it is well-known that one-loop RG coefficients for couplings with vanishing mass dimension are scheme-independent and universal. Furthermore, the two-loop gauge contribution to the gauge $\beta$-function is also known to be universal, provided a mass-scale independent regularisation scheme is adopted. Coefficients at higher loop order are strictly non-universal.  The main new effect in our work arises from the two-loop coefficients in the gauge sector, and from the interacting UV fixed point in the Yukawa RG flow at one-loop \eq{LOY}. Expressing the Yukawa fixed point in terms of the gauge coupling $\al y^*=\al y^*(\al g)$, one then shows that the fixed point in the gauge sector is invariant to leading order in $\eps$ under perturbative (non-singular) reparametrisations $\al g\to \al g'(\al g)$, see \eq{LOY}. We therefore conclude that the interacting UV fixed point arises universally, irrespective of the regularisation scheme.

\subsection{Operator ordering}
Unlike in asymptotically free theories, at an interacting UV fixed point  it is not known beforehand which invariants will become relevant since canonical power counting cannot be taken for granted  \cite{Falls:2013bv}. For asymptotically safe theories with perturbatively small  anomalous dimensions and corrections-to-scaling, however, canonical power counting can again be used to conclude that invariants with canonical mass dimension larger than four will remain irrelevant at a perturbative UV fixed point. The reason for this is that  corrections to scaling, in the regime \eq{small}, 
are too small to change canonical scaling dimensions by an integer, and hence cannot change irrelevant into relevant operators.  If masses are switched-on, two such operators are the fermion and scalar mass terms, both of which receive only perturbatively small corrections at the fixed point.  We conclude that the relevancy of operators continues to be controlled by their canonical mass dimension \cite{Falls:2013bv}.

\begin{figure*}[t]
\begin{center}
\includegraphics[width=.6\hsize]{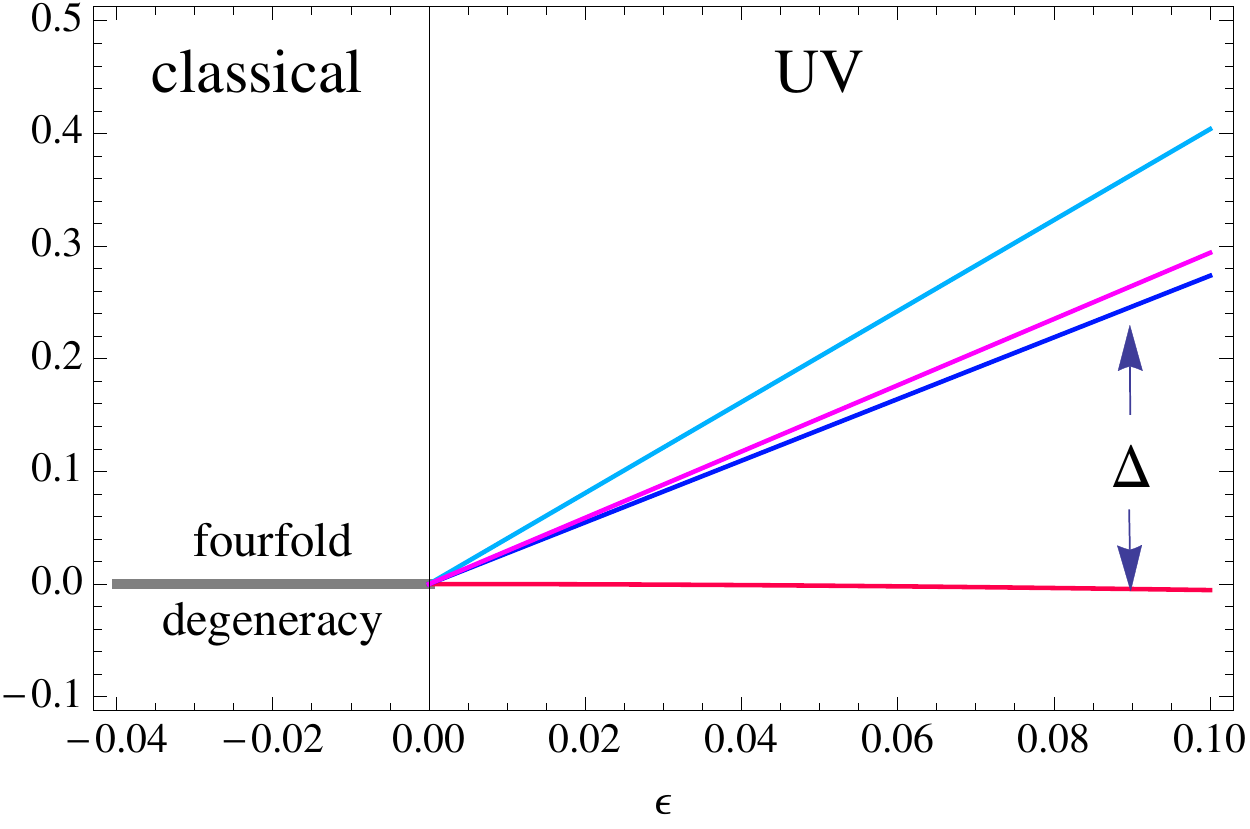}
\caption{\label{pDeltaAll} The fourfold degeneracy of the classically  marginal invariants \eq{F2}, \eq{Y}, \eq{U} and \eq{V} --- schematically indicated by 
a thick grey line (left panel) --- is lifted by residual interactions in the UV,  $0<\eps$  (right panel). Also shown are the universal  eigenvalues $\vartheta_1<0<\vartheta_2<\vartheta_4<\vartheta_3$  (bottom to top, respectively) of the fixed point  FP${}_1$, and the interaction-induced gaps $\Delta$ in the eigenvalue spectrum at NNLO as functions of $\eps$.}
 \vskip-.3cm
\label{pThetaAll}
\end{center}
\end{figure*}

On the other hand, residual interactions, even if perturbatively weak,  control the scaling of invariants which classically have a vanishing canonical mass dimension and can change these into relevant or irrelevant ones, see Fig.~\ref{pDeltaAll}. 
In our model, we find that the operator ordering of the classically marginal invariants at the fixed point is reflected by our search strategy, see Tab.~\ref{Tab}. 
At LO, the  $SU(N_C)$ Yang-Mills Lagrangean \eq{F2} coupled to $N_F$ fermions \eq{F} is assumed to become a relevant operator in the regime \eq{small} because asymptotic freedom is lost. This assumption is tested and confirmed at NLO against the inclusion of Yukawa interactions \eq{Y}. 
The eigenvalue $\vartheta_1$ is dominated by the gauge and $\vartheta_2$ dominated by the Yukawa coupling. This is consistent with the initial assumption inasmuch as the scaling of the Yukawa term provides a subleading correction to the scaling of the Yang-Mills term. 
At NNLO, two quartic scalar selfinteractions are introduced whose non-trivial fixed points add two eigenvalues to the spectrum. At FP${}_1$, both of these are irrelevant. At FP${}_2$, the double trace scalar selfinteraction becomes relevant. The structure of the scalar sector is induced by the fixed point in the gauge-Yukawa subsector. In general, for other values of the gauge and Yukawa couplings, the scalar sector may not offer a fixed point at all.

\subsection{Gap}
Residual interactions at the UV fixed point have lifted the fourfold degeneracy amongst the classically marginal couplings. In Fig.~\ref{pDeltaAll}, we show the eigenvalues to leading order in $\eps$ at the fixed point FP${}_1$, except for $\vartheta_1$ which is shown at order $\eps^2$. The difference between the smallest negative and the smallest positive eigenvalue, which we denote as the gap of the eigenvalue spectrum $\Delta$, is then a good quantitative measure for the strength of residual interactions.  At the UV fixed point we have $\Delta=\vartheta_2-\vartheta_1$. 
Quantitatively, the gap in the eigenvalue spectrum read
\beq
\begin{array}{rcl}
\Delta&= &
\displaystyle
\frac{52}{19}\eps+
{\cal O}(\eps^2)\,,
\end{array}
\eeq
where the (sub)leading term in $\eps$ arises from the (N)NLO approximation. Classically, we have $\Delta=0$. We notice that the leading and subleading term have the same sign, increasing the gap with increasing $\eps$. We stress that the gap in the eigenvalue spectrum is insensitive to the details of the scalar sector and only determined by the gauge-Yukawa subsystem.

\subsection{Unitarity}
An important constraint on quantum corrections relates to the scaling dimension of primary fields such as scalar fields themselves. For a quantum theory to be compatible with unitarity, it is required that the scaling dimension must be larger than unity, $\Delta_H>1$. This behaviour can be observed in the result. To leading order in $\eps$, $\gamma_H$ is negative and hence $\Delta_H>1$. At NNLO, we observe  cancellations in \eq{gammaH2} ensuring that $\gamma_H$ remains negative. Overall, fluctuation-induced corrections reach values of up to 5\% for moderate $\eps$. 

For the composite scalar operator $\delta^{ij}\bar Q_i Q_j$, the leading order corrections in $\eps$ decrease its scaling dimension $\Delta_F$ below its classical value $\Delta_F=3$, see \eq{adimensionF}. This is further decreased at  NNLO where all corrections to $\Delta_F$ have the same  sign and no cancellations occur. The NNLO corrections are thus stronger than those for $\Delta_H$. Here, corrections push $\Delta_F$ down from its classical value by up to 10\%, leaving $\Delta_F>1$. We conclude that the effects of residual interactions are compatible with basic constraints on the scaling of scalar operators.

\subsection{Triviality}\label{triviality}

Triviality bounds often  arise when infrared free interactions display a perturbative Landau pole towards high energies, limiting the predictivity of the theory to the scale of maximal UV extension \cite{Callaway:1988ya}.  On a more fundamental level, triviality relates to the  difficulty of defining a self-interacting scalar quantum field in four dimensions \cite{Wilson:1973jj,Luscher:1987ek,Hasenfratz:1987eh,Rosten:2008ts}, which also puts the existence of elementary scalars into question. In the standard model, the scalar and the $U(1)$ sectors are infrared free. 
At the UV fixed points detected here, triviality for all three types of fields is evaded through residual interactions. 
This also indicates that the scalar degrees of freedom may indeed be taken as elementary. 

Moreover, we also observe that the avoidance of triviality in the scalar sector is closely  linked to the presence of gauge fields, be they asymptotically free or asymptotically safe.
In fact, an interacting fixed point in the scalar sector would not arise without an interacting fixed point for the Yukawa coupling, see \eq{NLOH}, \eq{NLOV}. Furthermore, without gauge fields, the  fermion-boson subsystem does not generate an interacting UV fixed point, and couplings cannot reach the Gaussian fixed point in the UV.
With asymptotically free  gauge fields (say, for small $\eps<0$), the UV fixed point for the Yukawa  coupling remains the trivial one, see \eq{LOY},  \eq{NNLO}. A  detailed inspection  then shows that complete asymptotic freedom follows,
albeit under certain constraints on the parameters \cite{Harada:1994wy}.
With asymptotically safe  gauge fields (for small $\eps>0$), complete asymptotic safety is achieved at two interacting UV fixed points (see Tab.~\ref{Tab1}). We conclude that triviality is evaded in the large-$N$ limit with and without asymptotic freedom in the gauge sector, although the specific details differ.  Asymptotic freedom in the gauge sector had to be given up for the Yukawa and scalar sectors to develop interacting UV fixed points.

\section{Towards asymptotic safety at strong coupling}\label{LargeCoupling}
It would be useful to understand the existence or not of UV fixed points in non-Abelian gauge theories with matter and away from the regime where asymptotic safety is realised perturbatively  and $\eps$ is small. In this section, we indicate some directions towards larger $\eps$, with and without scalar matter.

\subsection{Beyond the Veneziano limit}\label{validity}

Presently, our study is bound to the second nontrivial order within perturbation theory and to the leading order in $1/N_F, 1/N_C\ll 1$, allowing for an accurate determination of the UV fixed point in the regime  \eq{small}. The stability in the result makes it conceivable that the UV fixed point may persist even for finite values of $\eps$.
With increasing $\eps$, the upper bound  \eq{max} which has arisen at NNLO comes into play. Solutions $(N_C,N_F)$ to the constraint 
\beq\label{solutionbound}
0\le\eps(N_C,N_F)<\eps_{\rm max}\,,
\eeq 
where we take for $\eps_{\rm max}$ its value at NNLO  given in \eq{max},
would then be likely candidate theories where the fixed point may exist even for finite but small couplings. The first few such solutions with the smallest numbers of fields are 
\beq
(N_C,N_F)=(5,28),(7,39),(9,50),(10,56),(11,61),(12,67),\cdots\,.
\eeq 
Once $N_C>12$,  more than one solution for $N_F$ may exist. 
Extending our study to N${}^3$LO should improve the estimate for the window \eq{solutionbound} for large $N$. For finite values of $N_F$ and $N_C$, the existence of an asymptotically safe window can in principle be tested using non-perturbative tools such as functional renormalisation \cite{Polchinski:1983gv,Wetterich:1992yh,Morris:1993qb,Litim:2001up}, or the lattice.

\subsection{Infinite order perturbation theory}
Interestingly, 
an infinite order result is available for 
nonabelian gauge theories  with a finite number of colors $N_C<\infty$, without scalars, but with $N_F\to \infty$  many Dirac fermions transforming according to a given representation of the gauge group \cite{Pica:2010mt}, see also \cite{Holdom:2010qs,Shrock:2013cca} and references therein. In the terminology of this work, this corresponds to the parameter regime
\beq\label{large}
1\ll \eps\,,
\eeq
see \eq{eps}. In this limit 
 the parametric deviation from asymptotic freedom is large, and the model becomes partly abelian  \cite{PalanquesMestre:1983zy}.
Defining $x = 4 N_F T_R\, \alpha$ with $\alpha = g^2/(4\pi)^2$ and $T_F$ the trace normalization,
 it is possible to sum exactly the infinite perturbative series for the gauge $\beta$-function for large numbers of flavors. The all-order result has the form  \cite{PalanquesMestre:1983zy,Gracey:1996he,Holdom:2010qs} 
\beq\label{zero}
\frac{3}{2x}
\frac{\beta(x)}{x} = 1 + \frac{H(x)}{N_F} + {\cal{O}}\left(N_F^{-2}\right)
\eeq
and an integral representation of $H(x)$ can be found in \cite{Holdom:2010qs} (see Fig.~\ref{pBetaNF} for an example). 
\begin{figure*}[t]
\begin{center}
\includegraphics[width=.6\hsize]{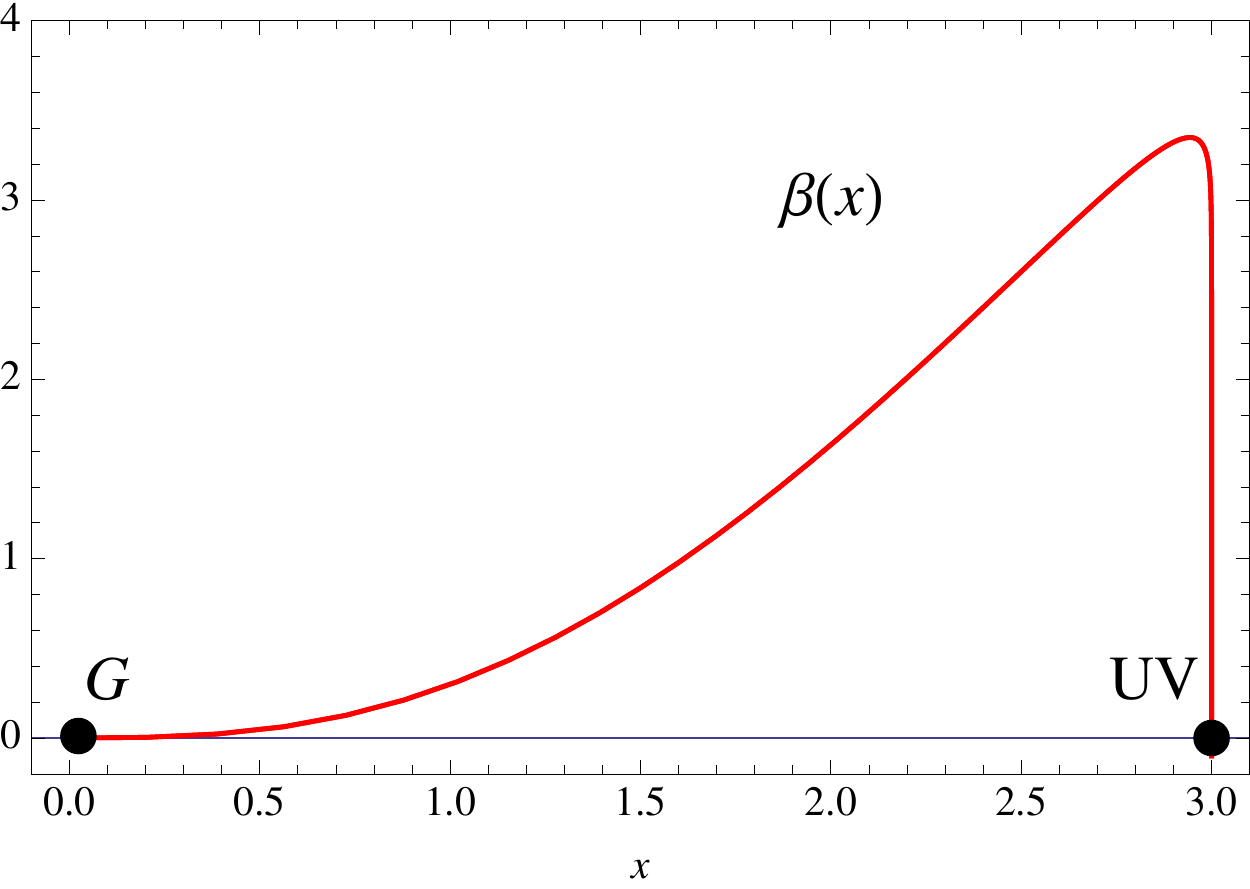}
\caption{\label{pBetaNF} The fully resummed gauge $\beta$ function \eq{zero} is shown to leading order in $1/N_F$ including the Gaussian and the UV fixed point ($N_C=5$ and $N_F=28$).}
 \vskip-.3cm
\label{pBetaNF}
\end{center}
\end{figure*} 
Adopting $T_F=\frac12$, one can show that the function $H(x)$ is finite for $0\le x<3$ with a  negative logarithmic singularity at $x= 3$ where $H(x)=N_C/8\times\ln|3-x|$ + const.~+${\cal O}(3-x)$. (Similar results are found for other representations as well.)
This structure implies the existence of a nontrivial UV fixed point to leading order in $1/N_F$. As a word of caution, however, we remind the reader that an infinite order perturbative result may be upset non-perturbatively, or by higher order terms in $N_F^{-1}$ 
(see \cite{Holdom:2010qs} for  a  discussion of the latter in QED). Expanding about the fixed point, the two leading terms read
\begin{equation}\label{no}
\alpha^{\ast} = \frac{3}{2\,N_F} 
-\frac1{2 N_F}\exp\left(-a\cdot \frac{N_F}{N_C}+b(N_C)\right)
\end{equation}
where $a=8$, and $b(N_C)\simeq 15.857+2.632/N_C^2$.
The UV fixed point starts dominating the RG running once 
$(\alpha^*-\alpha)/\alpha^*\klgl N_C/(16\,N_F)$ and its basin of attraction becomes algebraically small for large $N_F$; see Fig.~\ref{pBetaNF}.
Using the explicit form for $H(x)$  we also find the universal eigenvalue at the fixed point \eq{no} of the infinite order $\beta$-function \eq{zero},
\beq\label{evInf}
\vartheta=
-\frac 34\frac{N_C}{N_F}\,\exp \left(a\cdot \frac{N_F}{N_C}-b(N_C)\right)
\eeq
By construction, the result \eq{evInf} is valid in the limit $N_C/N_F\ll 1$ where  the eigenvalue becomes parametrically large.  
The exponent $\nu=-1/\vartheta$ for the correlation length becomes very small, $\nu\to 0$. Eigenvalues which grow rapidly with the number of degrees of freedom have been observed previously for quantum gravity in the large dimensional limit in the continuum \cite{Litim:2003vp,Fischer:2006fz,
Fischer:2006at} and from lattice considerations \cite{Hamber:2005vc}.

\subsection{Finite order perturbation theory}
The origin of asymptotic safety  in Yang-Mills theory with \eq{no}, \eq{evInf} is  different from the one observed in Sec.~\ref{AFAS}, because the vanishing of the gauge $\beta$-function \eq{zero} arises as an infinite order effect due to gluon and fermion loops for large $\eps$, rather than through an order-by-order cancellation of fluctuations from gauge, fermion and scalar fields for small $\eps$. It would be useful to understand whether the result \eq{no} persists beyond the limit of infinite $N_F$ with fixed and finite $N_C$. To that end, we test the continuity of the fixed point in $(N_F,N_C)$  by combining two observations. Firstly, we notice that a precursor of the fixed point \eq{no} is already visible within perturbation theory at finite orders. To see this  explicitly, we come back to our equations at NNLO and switch off the Yukawa and scalar coupling,  $\alpha_h=0,\alpha_y=0$ and $\al v=0$. In the parameter regime \eq{large}, we then find the  UV fixed point $\alpha^*_g={3}/(2\sqrt{7\eps})$ + subleading, and the eigenvalue $\vartheta=-4\sqrt{\eps}/\sqrt{7}$ + subleading.  Adopting the same definition for the coupling as in \eq{no}, this result translates into
\beq\label{inf}
\begin{array}{rcl}
\alpha^*&=&
\displaystyle
\frac{3}{2\sqrt{7}}\,\frac{1}{\sqrt{N^{}_C\,N_F}}
\,,\\[4ex]
\vartheta&=&
\displaystyle
-\frac{4}{\sqrt{7}}\,\sqrt{\frac{N_F}{N_C}}
\,,
\end{array}
\eeq
to leading order in $1/\eps$ and $1/N_F$. A few comments are in order. Comparing \eq{inf} with \eq{no}, \eq{evInf} 
for fixed $N_C$, we find that the $1/N_F$ decay of the fixed point is replaced by a softer square-root decay due to the finite order approximation in perturbation theory. The non-analytic dependence on $N_F$ and $N_C$ develops into the result \eq{no} with increasing orders in perturbation theory where the power law behaviour becomes $\alpha^*\sim N_F^{(2-p)/(p-1)}$ \cite{Pica:2010mt}, provided the $p$-loop coefficient is negative \cite{Shrock:2013cca}. We also find  that the eigenvalue $\vartheta$ in \eq{inf} 
grows  large in the regime \eq{large}, modulo subleading corrections. While the growth rate $\vartheta \sim -\sqrt{N_F}$ in \eq{inf} is weaker than the one observed in \eq{evInf}, the correlation length exponent $\nu$ shows the same  qualitative behaviour $\nu\to 0$ as the infinite order fixed point.   
We thus may conclude that \eq{inf}  is the low-order precursor to the all-order result \eq{no}. 

Secondly, for the finite order fixed point \eq{inf} we observe that the limits $1/N_C\to 0$ with $N_F/N_C$ fixed can be accessed, and hence  finite values for $\eps$ with \eq{large}, because the underlying NNLO equations remain valid in this parameter regime.  Note that this limit is not covered by the rationale which has led to \eq{zero}. The UV fixed point then reads
\beq\label{no1}
\alpha^*=\frac{33+6\eps}{4\sqrt{7\,\eps}}\frac{1}{N_F}\,.
\eeq
For fixed $\eps$, the fixed point  shows the same $1/N_F$ behaviour as the fixed point \eq{no}. The coefficient in front of $1/N_F$ in \eq{no1} is larger than the fixed point \eq{no} for all finite $\eps$. 
Unlike \eq{evInf}, its eigenvalue \eq{inf} remains bounded since $N_F/N_C$ is finite. It would thus seem that the inclusion of more gluons or less fermions maintains the UV fixed point, albeit with a softened UV scaling behaviour and at stronger coupling. 
The continuity of results in $(N_F,N_C)$ suggests that the UV fixed point \eq{no} is not an artefact of the  large-$N_F$ limit, but rather a fingerprint of a fixed point in the physical theory.

In summary, the observations in this section indicate that the
matter-gauge systems studied here have a sufficiently rich structure to admit asymptotically safe UV fixed points also for finite $(N_C,N_F)$, with and without scalar matter, in addition to the weakly coupled UV fixed point for small $\eps$. More work is required to identify them reliably within perturbation theory and beyond, and for generic values of $\eps$. 

\section{Conclusion}\label{discussion}

We have used large-$N$ techniques to understand the ultraviolet behaviour of theories
involving fundamental gauge fields, fermions, and scalars. In strictly four space-time dimensions, and  in the regime where the gauge sector is no longer asymptotically free, we have identified a perturbative origin for asymptotic safety. 
We  found
that all three types of fields are necessary for an interacting UV fixed point to arise.
The primary driver towards asymptotic safety are the Yukawa interactions, which source the interacting fixed point for both the gauge fields and the scalars. In return, the gauge fields stabilise an interacting fixed point in the Yukawa sector.
Fixed points are established in the perturbative domain, consistent with unitarity.
Triviality bounds and Landau poles are evaded. Here the scalar fields can  be considered as elementary.

It would be worth extending this picture within perturbation theory and beyond, also taking subleading corrections 
into consideration, and for fields with more general gauge charges, gauge groups, and Yukawa interactions. Once the number of fields is finite, asymptotic safety can in principle be tested non-perturbatively using the powerful machinery of functional renormalisation \cite{Polchinski:1983gv,Wetterich:1992yh,Morris:1993qb,Litim:2001up}, or the lattice.  In a different vein, one might  wonder whether the weakly coupled ultraviolet fixed point has a strongly coupled dual. First steps  to extend the ideas of Seiberg duality \cite{Seiberg:1994pq} to  non-supersymmetric theories have been discussed in \cite{Sannino:2009qc,Sannino:2010fh}. It has also been suggested that UV conformal matter could simplify the quantisation of canonical gravity \cite{Hooft:2010ac}, or help to resolve outstanding puzzles in particle physics and cosmology. Our study offers such candidates.

\acknowledgments

This work is supported by the Science Technology and Facilities Council (STFC) [grant number ST/J000477/1], by the National Science Foundation under Grant No.~PHYS-1066293, and by the hospitality of the Aspen Center for Physics. The CP${}^3$-Origins centre is partially funded by the Danish National Research Foundation, grant number DNRF90.

\bibliographystyle{apsrev4-1}
\bibliography{final}

\end{document}